Friedrich-Schiller-Universität Jena

seit 1558

## Studienarbeit

# Fixed-Parameter Algorithms for Computing Kemeny Scores – Theory and Practice

ausgeführt am Lehrstuhl für Theoretische Informatik I
an der Falkultät für Mathematik und Informatik der
Friedrich-Schiller-Universität Jena

unter Anleitung von Dipl.-Bioinf. Nadja Betzler und Prof. Dr. Rolf Niedermeier

durch

Robert Bredereck
Mat.Nr.: 84906

Oktober 2009

# Erklärung

Hiermit erkläre ich, dass ich die Arbeit selbständig und nur mit den angegebenen Hilfsmitteln angefertigt habe.

Jena, 14. Oktober 2009

# Abstract

The central problem in this work is to compute a ranking of a set of elements which is “closest to” a given set of input rankings of the elements. We define “closest to” in an established way as having the minimum sum of Kendall-Tau distances to each input ranking. Unfortunately, the resulting problem Kemeny consensus is NP-hard for instances with $n$ input rankings, $n$ being an even integer greater than three. Nevertheless this problem plays a central role in many rank aggregation problems. It was shown that one can compute the corresponding *Kemeny consensus list* in $f(k)+\text{poly}(n)$ time, being $f(k)$ a computable function in one of the parameters “score of the consensus”, “maximum distance between two input rankings”, “number of candidates” and “average pairwise Kendall-Tau distance” and $\text{poly}(n)$ a polynomial in the input size. This work will demonstrate the practical usefulness of the corresponding algorithms by applying them to randomly generated and several real-world data. Thus, we show that these fixed-parameter algorithms are not only of theoretical interest. In a more theoretical part of this work we will develop an improved fixed-parameter algorithm[1] for the parameter “score of the consensus” having a better upper bound for the running time than previous algorithms.

[1]Independently from this work, an even more improved algorithm was developed in [Sim09].

# Contents

# 1. Introduction

## 1.1. Kemeny's voting scheme

There are many situations, where one has to get an ordered list of *candidates* by aggregating inconsistent information. For example, in plurality voting systems each voter determines which candidate is the best. He[1] cannot affect the order of the remaining candidates among each other. Our aim is to get an order of the candidates that best reflects the opinion of the voters. The disadvantage is that the information (which is provided or used) of each vote is incomplete in respect to the solution. Of course, there are also some advantages: Sometimes it might be easier for a voter to determine his vote because he only has to know who is the best for him. There are efficient ways to compute the resulting preference list. To analyse attributes of different (and more complex) voting systems, we introduce a formal view of a voting system. The input is an *election* $(V, C)$ consisting of a set $V = \{v_1, v_2, ..., v_n\}$ of votes over a set $C$ of $m$ candidates. One vote is a *preference list* of the candidates, that is, each vote puts the candidates in an ordered list according to preference. The solution is a single preference list, whose computation depends on the respective voting system. Although we can use this formalism already for plurality voting systems, there are many situations with more intricate voting scenarios. For example, different sports competitions lead to a voting scenario, where we have preference lists anyway. For instance, the results of each race in one Formula One season form inconsistent information about the skills of the drivers. At the end of each season we do not only want to see a world champion, but also a complete preference list of drivers refering to their skills. The FIA[2] has used several point-scoring systems [Wik09b] to determine the overall preference list. None of these systems took the whole race results into account. As a consequence, the overall result might not fairly reflect the driver's skills.

**Example 1**
In a fictive season there are the two drivers Adrian and Bob and 14 other drivers. We use the point-scoring system of the year 2003 till present (2009):

| | | | |
|---|---|---|---|
| 1st place | 10 points | 5th place | 4 points |
| 2nd place | 8 points | 6th place | 3 points |
| 3th place | 6 points | 7th place | 2 points |
| 4th place | 5 points | 8th place | 1 point |

[1] For the sake of simplicity we use male sex for all candidates. This also applies to drivers, politicians, and so on.
[2] Fédération Internationale de l'Automobile

At the end of the season, the drivers are ranked according the point sums. Adrian is the last driver who passes the finish line in each race. In one race eight drivers (inclusive Bob) fail so that Adrian gets one point. In all other races, Bob is getting the 9th place and no points. Finally the point-scoring system ranks Adrian better than Bob while it is obvious that Bob was "more successful" in that season.

Although this example is overstated, it illustrates the problem of using a voting scenario that only uses a (small) subset of the pairwise relations between the candidates. Thus, it is desirable to use a voting system that reflects the whole race results. In this case "reflecting the whole input information" means, that each position in the preference list of a vote may affect the solution list. (It is obvious, that the plurality voting system does not reflect the whole input information.) Borda is a well-known example among point-scoring systems. The *Borda* (or Borda count) voting system determines the winner of an election by giving each candidate a certain number of points corresponding to the position in which he is ranked by each voter. As result, we get a preference list, where all candidate are ranked according to their points sums. Furthermore, we take another important attribute of voting systems into account. Informally, the Condorcet winner[3] is the candidate who would win a two-candidate election against each of the other candidates (Definition 1). Unfortunately, there is no guarantee that the Borda winner is also a the Condorcet winner [Kla05].

**Definition 1.** *The **Condorcet winner** of an election is the candidate who, when compared with each other candidate, is preferred to every other candidate in more than half of the votes. A voting system satisfies the **Condorcet criterion** if it chooses the Condorcet winner when one exists.*

**Example 2**

Consider the election $(V, C)$ with $V = \{v_1, \ldots, v_5\}$ and $C = \{a, b, c\}$. Each voter assigns three points for the most preferred candidate, two points for the secondary most preferred candidate and one point fast the least preferred candidate. We have the following votes:

$$\begin{aligned} v_1 :&\quad a > b > c \\ v_2 :&\quad a > b > c \\ v_3 :&\quad a > b > c \\ v_4 :&\quad b > c > a \\ v_5 :&\quad b > c > a \end{aligned}$$

In other words, $a$ gets 11 points, $b$ gets 12 points and $c$ gets 7 points. The Borda winner is $b$ although the Condorcet winner $a$ is in three of five votes better than each other candidate.

[3] A Condorcet winner will not always exist in a given set of votes, which is known as Condorcet's voting paradox.

One famous voting system that satisfies the Condorcet criterion is *Kemeny's voting scheme*. It goes back to Kemeny [Kem59] and was specified by Levenglick [Lev75] in 1975. The result of this voting scheme is the so-called *Kemeny consensus*. It is a preference list $l$, that is "closest" to the input preference lists of the votes. In this case "closest" is formally defined as the minimum sum of *Kendall-Tau distances* between $l$ and each vote $v_i$. The Kendall-Tau distance between the votes $v$ and $w$ is defined as

$$\text{KT-dist}(v, w) = \sum_{\{c,d\} \subseteq C} d_{v,w}(c, d) \qquad (1.1)$$

where the sum is taken over all unordered pairs $\{c, d\}$ of candidates, and $d_{v,w}(c, d)$ is set to 0 if $v$ and $w$ rank $c$ and $d$ in the same relative order, and otherwise it is set to 1. Using a divide-and-conquer algorithm, one can compute the Kendall-Tau distance in $O(m \cdot \log m)$ [KT06]. We define the *score* of a preference list $l$ in an election $(V, C)$ as $\sum_{v \in V} \text{KT-dist}(l, v)$. That is, the Kemeny consensus (or Kemeny ranking) of $(V, C)$ is a preference list with minimum score, called the *Kemeny score* of $(V, C)$. Clearly, there can be more than one optimal preference list. Altogether, we arrive at the decision problem behind the computation of the Kemeny consensus:

Kemeny Score
*Input:* An election $(V, C)$ and a positive integer $k$.
*Output:* Is the Kemeny score of $(V, C)$ at most $k$?

All algorithms in this work do not only solve Kemeny Score itself, but also compute the optimal score and a corresponding consensus list for the given election.

While using sports competition results to define input preference lists is easy, it seems more difficult to use Kemeny's voting scheme for voting systems with many candidates. The voters may not be able or not disposed to provide a complete preference list for all candidates. An example is, when four persons of the human resources department have to determine a ranking of hundred applicants. Here, the goal might be to select the top five applicants and each human resources person provides a ranking of all applicants. Of course, there are also special situations with only a few candidates where the voters provide complete preference lists. In case of local elections in German politics we usually have only five till ten candidates. (Nevertheless, a majority voting system is used for these candidates.) However, in politics voting systems that use preference lists as input are very rare at present. They are for example used to elect members of the Australian House of Representatives, the President of Ireland, the national parliament of Papua New Guinea, and the Fijian House of Representatives [Wik09a]. There are many other scenarios where it is easy to extract a set of preference lists from the input information. For example, Kemeny's voting scheme is used in genetic analysis [JSA08], meta search engines [DKNS01a], database applications [FKM+04], or fighting spam [LZ05, CDN05]. Therefore the performance of solving Kemeny Score is important. In the following paragraph we will summarize the state of the art regarding the classical computational complexity of Kemeny Score.

**Complexity** Bartholdi et al. [BTT89] showed that KEMENY SCORE is NP-complete. Since KEMENY SCORE has practical relevance, polynomial-time algorithms are highly desirable. So there are several studies for approximation algorithms with polynomial running time. A deterministic approximation algorithm with factor 8/5 was shown by van Zuylen et al. [vZW07]. With a randomized algorithm it is possible to improve the factor to 11/7 [ACN08]. Recent studies [KMS07] showed that there is also a polynomial-time approximation scheme (PTAS) for KEMENY SCORE, but the corresponding running time is not practical. In several applications exact solutions are indispensable. Hence, a parameterized complexity analysis might be a way out. That is why we concentrate on methods of parameterized algorithms in the following. The next paragraph contains a survey of our results.

**Survey** In this work we will analyse and develop algorithms that solve KEMENY SCORE efficiently when the parameter $k$ being the Kemeny score is small. More precisely we will provide an algorithm that decides KEMENY SCORE in $O(1.5079^k + m^2 \cdot n)$ time. This is an improvement from $O(1.53^k + m^2 \cdot n)$ in previous work [BFG+09]. We will discuss some tricks and heuristics to improve the running time in practice and develop a polynomial-time data reduction rule. Together with an implementation of another algorithm in [BFG+09] solving KEMENY SCORE efficiently when the parameter "number of candidates" is small, we will get a framework to compute optimal solutions for real-world instances with up to 30 candidates. (The number of votes do not affect the running-time noticeable.) We will show that we can use Kemeny rankings to evaluate sports competitions and to create meta search engines without counting on heuristics and approximative solutions. Tests on real-world data will show that data reduction rule seems to be useful.

## 1.2. Preliminaries

Some basic definitions were already given in Section 1.1. Now, we will define further terms that are fundamental for the next sections. Let the *position* of a candidate $a$ in a vote $v$ be the number of candidates who are better than $a$ in $v$. Thus, the best (leftmost) candidate in $v$ has position 0 and the rightmost candidate has position $m-1$. Then $\text{pos}_v(a)$ denotes the position of candidate $a$ in $v$.

**Definition 2.** *Let $(V, C)$ be an election. Two candidates $a, b \in C, a \neq b$, form a* ***dirty pair*** *if there exists one vote in $V$ with $a > b$ and there exists another vote in $V$ with $b > a$. A candidate is called* ***dirty*** *if he is part of a dirty pair, otherwise he is called* ***non-dirty***.

This definition is very important for the next sections. Later we will extend this concept of "dirtiness" to analyse the complexity of an algorithm. We illustrate the definition by Example 3.

**Example 3**
We have an election $(V, C)$ with $V = \{v_1, v_2, v_3\}$ and $C = \{a, b, c, d, y\}$.

$$v_1: \quad a > b > y > c > d$$
$$v_2: \quad b > a > y > c > d$$
$$v_3: \quad a > b > y > d > c$$

The relative orders of the pairs $\{a, c\}$, $\{a, d\}$, $\{b, c\}$, $\{b, d\}$ and $\{x, y\}$ for $x \in \{a, b, c, d\}$ are the same in all votes, but there is at least one vote for each possible relative order of $\{a, b\}$ and $\{c, d\}$. Thus, we have two dirty pairs $\{a, b\}$ and $\{c, d\}$ and one non-dirty candidate $y$. All other candidates are dirty.

In this work the terms "preference list of candidates" and "permutation of candidates" are used equivalently. This means, for example, that the preference list $a > b > c > d$ is equivalent to the permutation $(a, b, c, d)$. Later on will will analyse algorithms that fix the relative order of some candidates. We have to consider that not all combinations of fixed relative orders are consistent. An example for an inconsistent combination of pairwise relative orders is as follows:

**Example 4**
Take three candidates $a$, $b$, and $c$, where each pair is dirty. A consensus can not have $a > b$, $b > c$ and $c > a$, because $a > b$ and $b > c$ implies $a > c$.

For the purpose of analysis we introduce a concept of *consistence* for a set of ordered pairs:

**Definition 3.** *Let $(V, C)$ denote an election and let $O$ denote a set of ordered pairs of candidates in $C$. Furthermore, let $p$ denote a preference list of the candidates in $C$. We say $O$ and $p$* ***agree*** *if $x > y$ in $p$ for each $(x, y) \in O$. If there exists a preference list $p$ and $O$ agrees with $p$, we call $O$* ***consistent****. We call $O$ the* ***relation set*** *of $p$ if $O$ agrees with $p$ and for each pair of candidates $\{x, y\}$ either $(x, y)$ or $(y, x)$ is in $O$. Finally, let $X$ and $Y$ denote two sets of ordered pairs. We say $X$ and $Y$* ***agree****, if there is no ordered pair $(k, l)$ in $X$ with $(l, k)$ in $Y$.*

Of course, the relation set of each preference list is uniquely determined.

**Example 5**
If we transfer the relations from Example 4 to a relation set we get the inconsistent set $O_1 := \{(a, b), (b, c), (c, a)\}$. Otherwise, the subset $O_2 := \{(a, b), (b, c)\}$ is consistent. It agrees for example with $p := a > b > c$. The relation set of $p$ is $O_3 := \{(a, b), (b, c), (a, c)\}$. Trivially $O_1$ and $O_2$ agree as well as $O_2$ and $O_3$, but $O_3$ and $O_1$ do not agree.

**Observation 1.** *Let $X$ and $Y$ denote two sets of ordered pairs. If $X$ and $Y$ do not agree, then $X \cup Y$ is not consistent.*

For later analysis purposes, we define the concept of the *subscore* of a set $O$ of ordered pairs of candidates for an election $(V, C)$:

$$\text{subscore}(O) = \sum_{v \in V} \sum_{\{c,d\} \subseteq C} d_v(c, d) \tag{1.2}$$

where $d_v$ is set 1 if the relative order of $c$ and $d$ in $v$ is not an element of the set $O$, and else $d_v$ is set 0. The following observation is trivial:

**Observation 2.** *Let $(V, C)$ be an election, let $p$ denote a preference list of candidates from $C$, and let $P = \{(x, y) \mid \text{pos}_p(x) < \text{pos}_p(y)\}$. Then,* $\text{subscore}(P' \subseteq P) \leq \text{score}(p)$.

It follows from Observation 2 that one can use the subscore to estimate the score of a preference list. In the following, we will use this observation to discard some branching cases and improve the running time in practice.

## 1.3. Fixed-parameter tractability

Many interesting problems in computer science are computationally hard problems. The most famous class of such hard problems like KEMENY SCORE is the class of NP-hard problems. The relation between P (which includes the "efficient solvable problems") and NP is not completely clear at the moment[4]. Even if P = NP it is not self-evident that we are able to design *efficient* polynomial-time algorithms for each NP-hard problem. But we have to solve NP-hard problems in practice. Thus, according to the state of the art of computational complexity theory, NP-hardness means that we only have algorithms with exponential running times to solve the corresponding problems exactly. This is a huge barrier for practical applications. There are different ways to cope with this situation: heuristic methods, randomized algorithms, average-case analysis (instead of worst-case) and approximation algorithms. Unfortunately, none of these methods provides an algorithm that computes an optimal solution in polynomial time in the worst case. Since there are situations where we need both, another way out is needed. Fixed-parameter algorithms provide a possibility to redefine problems with several input parameters. The main idea is to analyse the input structure to find parameters that are "responsible for the exponential running time". The aim is to find such a parameter, whose values are constant or "logarithmic in the input size" or "usually small enough" in the problem instances of your application. Thus, we can say something like "if the parameter is small, we can solve our problem instances efficiently".
We will use the two dimensional parameterized complexity theory [DF99, Nie06, FG06] for studying the computational complexity of KEMENY SCORE. A *parameterized problem* (or language) $L$ is a subset $L \subseteq \Sigma^* \times \Sigma^*$ for some finite alphabet $\Sigma$. For an element $(x, k)$ of $L$, by convention $x$ is called *problem instance*[5] and $k$ is the *parameter*. The two dimensions of parameterized complexity theory are the size of the input $n := |(x, k)|$ and the parameter value $k$, which is usually a non-negative integer. A parameterized language is called *fixed-parameter tractable* if we can determine in $f(k) \cdot n^{O(1)}$ time whether $(x, k)$ is an element of our language, where $f$ is a computable function only depending on the parameter $k$. The class of fixed-parameter tractable problems is called FPT. Summarizing, the intention of parameterized complexity theory is to confine the combinatorial explosion to the parameter. The parameter can be nearly anything so that not all parameters are very helpful. Thus, it is very important to find good parameters.

[4] G. Woeginger maintains a collection of scientific papers that try to settle the "P versus NP" question (in either way) [Woe09]

[5] Most parameterized problems originate from classical complexity problems. You can see $x$ as the input of the original/non-parameterized problem.

In the following sections, we need two of the core tools in the development of parameterized algorithms [Nie06]: data reduction rules (kernelization) and search trees. The idea of *kernelization* is to transform any problem instance $x$ with parameter $k$ in polynomial time into a new instance $x'$ with parameter $k'$ such that the size of $x'$ is bounded from above by some function only depending on $k$ and $k' \leq k$, and $(x, k) \in L$ if and only if $(x', k') \in L$. The reduced instance $(x', k')$ is called *problem kernel*. This is done by *data reduction rules*, which are transformations from one problem instance to another. A data reduction rule that transforms $(x, k)$ to $(x', k')$ is called *sound* if $(x, k) \in L$ if and only if $(x', k') \in L$.

Besides kernelization we use (depth-bounded) *search trees algorithms*. A search algorithm takes a problem as input and returns a solution to the problem after evaluating a number of possible solutions. The set of all possible solutions is called the search space. Depth-bounded search tree algorithms organize the systematic and exhaustive exploration of the search place in a tree-like manner. Let $(x, k)$ denote the instance of a parameterized problem. The search tree algorithm replaces $(x, k)$ by a set $H$ of smaller instances $(x_i, k_i)$ with $|x_i| < |x|$ and $k_i < k$ for $1 \leq i \leq |H|$. If a reduced instance $(x', k')$ does not satisfy one of the *termination conditions*, the algorithms recursively applies the replacing procedure to $(x', k')$. The algorithms terminates if at least one termination condition is satisfied or the replacing procedure is no longer applicable. Each recursive call is represented by a search tree node. The number of search tree nodes is governed by linear recurrences with constant coefficients. There are established methods to solve these recurrences [Nie06]. When the algorithm solves a problem instance of size $s$, it calls itself to solve problem instances of sizes $s - d_1, \ldots, s - d_i$ for $i$ recursive calls. We call $(d_1, \ldots, d_i)$ the *branching-vector* of this recursion. So, we have the recurrence $T_s = T_{s-d_1} + \cdots + T_{s-d_i}$ for the asymptotic size $T_s$ of the overall search tree. The roots of the characteristic polynomial $z^s = z^{s-d_1} + \cdots + z^{s-d_i}$ with $d = max\{d_1, \ldots, d_i\}$ determine the solution of the recurrence relation. In out context, the characteristic polynomial has always a single root $\alpha$, which has the maximum absolute value. With respect to the branching vector, $|\alpha|$ is called the *branching number*. In the next chapter, we will analyse search tree algorithms that solve KEMENY SCORE.

# 2. Fixed-parameter tractability of the Kemeny score problem

In this chapter we will analyse known parameterized algorithms for the KEMENY SCORE problem. We will examine a search tree algorithm from [BFG+08] and extend it by a more refined branching strategy, which depends on a special program parameter $s$ being the "size of the branching object". We will analyse its worst-case running time for $s = 4$. Although there is hope that the worst-case running time is even better for greater values, we will see that our search tree algorithm is better than the previous one in [BFG+08]. In addition, we give a data reduction rule. Its exhaustive application also solves a special case in polynomial time. We will start with an overview of recent studies of the KEMENY SCORE problem in parameterized manner.

## 2.1. Known results

We already know that KEMENY SCORE is NP-hard. At present, this means that computing an optimal Kemeny consensus takes exponential time in worst case. In several applications, exact solutions are indispensable. Hence, a parameterized complexity analysis might be a way out. Here one typically faces an exponential running time component depending only on a certain parameter, cf. Section 1.3. An important parameter for many problems is the size of the solution. In case of KEMENY SCORE this is the "score of the consensus". Betzler et al. [BFG+08] showed that KEMENY SCORE can be solved in $O(1.53^k + m^2 \cdot n)$ time with $k$ being the score of the consensus. They also showed that one can solve KEMENY SCORE in $O((3d + 1)! \cdot d \log d \cdot mn)$ time with $d$ being the maximum Kendall-Tau distance between two input votes, in $O(2^m \cdot m^2 \cdot n)$ time and in $O((3r+1)! \cdot r \log r \cdot mn)$ time with $r$ being the maximum range of candidate positions. Another interesting parameter for parameterized computational complexity analysis is of course "number of votes", but Dwork et al. [DKNS01a, DKNS01b] showed that the NP-completeness holds even if the number of votes is only four. Hence, there is no hope for fixed-parameter tractability with respect to this parameter. In recent studies, Betzler et al. [BFG+09] showed that KEMENY SCORE can be solved in $O(n^2 \cdot m \log m + 16^{d_A} \cdot (16{d_A}^2 \cdot m + 4d_A \cdot m^2 \log m \cdot n))$ time with $d_A = \lceil d_a \rceil$ and $d_a$ being the average Kendall-Tau distance. Furthermore, this is clearly an improved algorithm for the parameterization by the maximum Kendall-Tau distance. Because the maximum range of candidate positions is at most $2 \cdot d_a$ [BFG+09], we also have an improved algorithm for the parameterization by the maximum range of candidate positions. In the next subsection, we will examine more closely the parameterized algorithms for the parameter "score of the consensus" [BFG+08]. Later on, we will improve the results and describe an algorithm that solves Kemeny Score in $O(1.5078^k + m^2 \cdot n)$. Independently from this work, an even more improved algorithm was developed in [Sim09]. They use the minimum-weight feedback arc set to provide a quiet similar branching strategy and to get an upper bound for the search tree size in $O(1.403^k)$.

### 2.1.1. Known results for the parameter Kemeny score

A trivial search tree for Kemeny Score can be obtained by branching on the dirty pairs. More precisely, we can branch into the two possible relative orders of a dirty pair at each search tree node. The parameter will be decreased at least by one in both cases. Actually, it will be decreased by more than one in many cases. Thus, we get a search tree of size $O(2^k)$. Since we want to compute the consensus list we also want to know the relative order of the non-dirty pairs. Fortunately, the relative order of all non-dirty candidates and all non-dirty pairs is already fixed:

**Lemma 1.** *Let $(V, C)$ be an election and let $a$ and $b$ be two candidates in $C$. If $a > b$ in all votes, then every Kemeny consensus has $a > b$.*

The correctness of Lemma 1 follows from the Extended Condorcet criterion [Tru98]. The fact of the following lemma is well-known. For the sake of completeness we provide a proof.

**Lemma 2.** Kemeny Score *is solvable in polynomial time for instances with at most two votes.*

*Proof.* For instances with one vote: Take the vote as consensus; its score is zero. For instances with two votes: Take one of the votes as consensus. The score will be $s_v := \sum_{\{a,b\} \subseteq C} d_{v_1,v_2}(a, b)$. For each preference list $p$ with $v_1 \neq p \neq v_2$ the score will be at least $s_v$, because for each pair $\{a, b\}$ it holds that $\sum_{v \in \{v_1,v_2\}} d_{p,v}(a, b) \geq d_{v_1,v_2}(a, b)$. This can be proved by contradiction: Assume that $\sum_{v \in \{v_1,v_2\}} d_{p,v}(a, b) < d_{v_1,v_2}(a, b)$. In this case $d_{v_1,v_2}(a, b)$ has to be 1. We show that $\sum_{v \in \{v_1,v_2\}} d_{p,v}(a, b)$ can not be 0. Since $v_1$ and $v_2$ rank $a$ and $b$ in a different order, $p$ and $v_1$ can not rank $a$ and $b$ in the same order if $p$ and $v_2$ rank $a$ and $b$ in the same order. □

It follows from Lemma 2 that we are only interested in instances with at least three votes. Let $\{a, b\}$ denote a dirty pair. The hardest case for the analysis of the branching number is if $a > b$ in only one vote. Then, there are at least two votes with $b > a$. This will help to do a better estimate of the branching vector in the search tree. As a consequence, having a look at the search tree again, we see that we can decrease the parameter by 2 in at least one of the two cases. Thus, it is easy to verify that the search tree size of this trivial algorithm is $O(1.618034^k)$. Betzler et al. [BFG+08] showed that there is an improved search tree by branching on *dirty triples*, that is a set of three candidates, such that at least two pairs of them are dirty pairs. The size of the resulting search tree is $O(1.53^k)$. Intuitively, there is hope that branching on a "dirty set with more than three candidates" will decrease the size of the search tree further. This is what we examine next.

### 2.1.2. A closer look on the search trees

Now, we closely examine the search tree algorithms that decide Kemeny Score as described in [BFG+08]. In the both search trees one computes a consensus list by fixing the relative order of the candidates of each dirty pair in one search tree node. In the *trivial search tree* we fix the relative order of the candidate of one dirty pair per search tree node. In the *triple search tree* we fix the relative orders of the candidates of all dirty pairs that are involved in one dirty triple per search tree node. At the node where we fix the order and at all child nodes of this node, we denote these dirty pairs as *non-ambiguous*. Intuitively, a pair is called *ambiguous* if the relative order of its candidates was not fixed. At every search tree leaf, all pairs are non-ambiguous so that the relative order of the candidates of each dirty pair is fixed. That is, the consensus list is uniquely determined if the fixed orders are consistent. At this point, we can make some observations:

**Observation 3.** *At each search tree node, the parameter decreases according to the subscore of the set of fixed pairs.*

We can compute the Kemeny score by summing up the subscores of the sets of fixed pairs. Clearly, each dirty pair will be fixed only once. Thus, Observation 3 is correct.

**Observation 4.** *One has the termination condition: If the set of non-ambiguous pairs is inconsistent, then discard the branching.*

Let $U$ denote the set of non-ambiguous pairs in a search tree node $u$. Then, $U$ is clearly a subset of the set of non-ambiguous pairs in each subtree node of $u$. Clearly, a superset of an inconsistent set is inconsistent, too. Thus, Observation 4 is correct.
The improvement in the triple search tree uses the following observation: In the search tree it does not affect the correctness in which sequence the dirty pairs are processed. In the trivial search tree we process the dirty pairs in arbitrary sequence. For the triple search tree, we can assume the we process all dirty pairs, involved in the same dirty triple, successively. We replace all search tree nodes that handle dirty pairs of the same dirty triple, with one new node, where we branch on all possible relative orders of the candidates of the dirty triple (see Figure 2.1 "Improving the search tree"). This lead to a decreased branching number. Our aim is to generalize this idea to a "dirty sets of arbitrary size" and get a more refined search tree algorithm. Observations 3 and 4 are valid for every branching strategy that fixes the relative order of candidates of each dirty pairs.

## 2.2. Refinement of the search tree

Now we want to design the more refined search tree. So, we need a concept of a structure of arbitrary size that extends the known terms "dirty pair" and "dirty triple".

We consider the search tree of the $O(2^k)$ algorithm. The following trees are only small sections of a complete search-tree for an election with at least five candidates $\{a, b, c, x, y\}$, where at least $\{a, b\}$, $\{b, c\}$, $\{a, c\}$ and $\{x, y\}$ are dirty pairs.

original tree - showing how relations get fixed

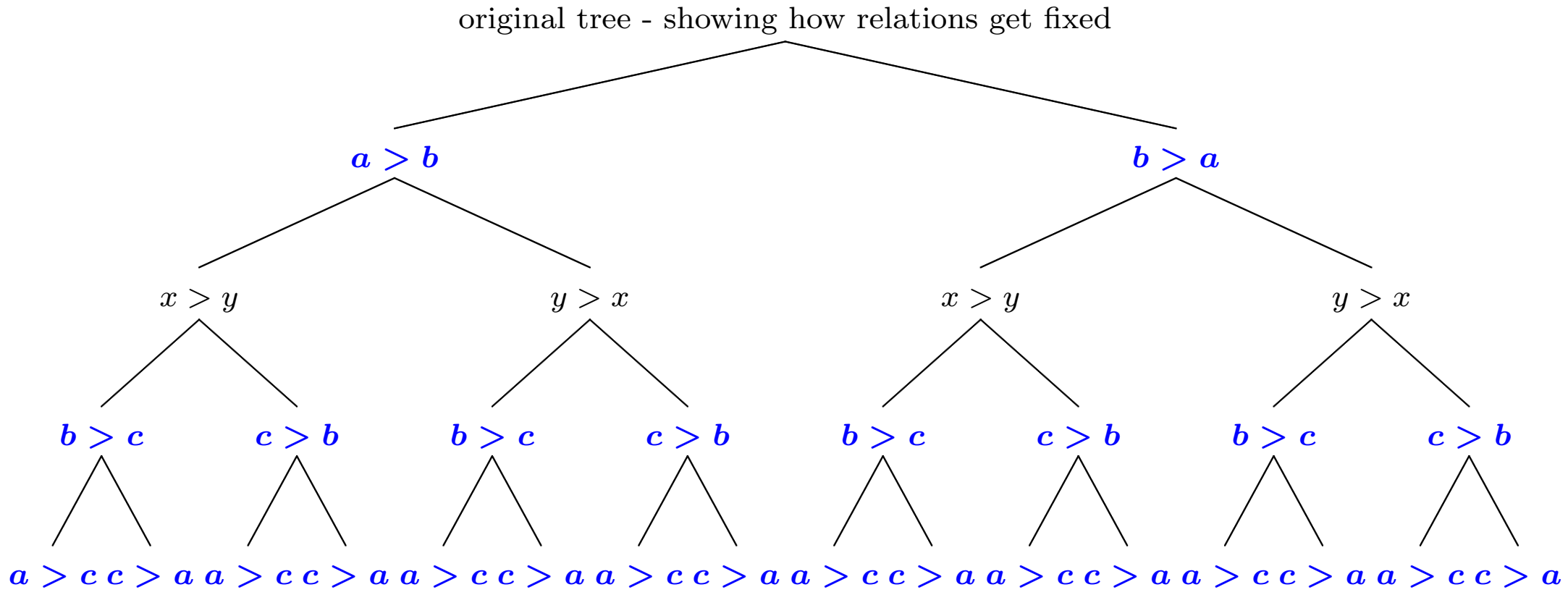

We will change the fixing order. Remark: The leaves of the new tree contain the same combination (of fixed relative orders) as the leaves of the original tree. Thus, a changed fixing order does not affect the correctness of the algorithm (completeness of the search-tree).

sorted pair sequence

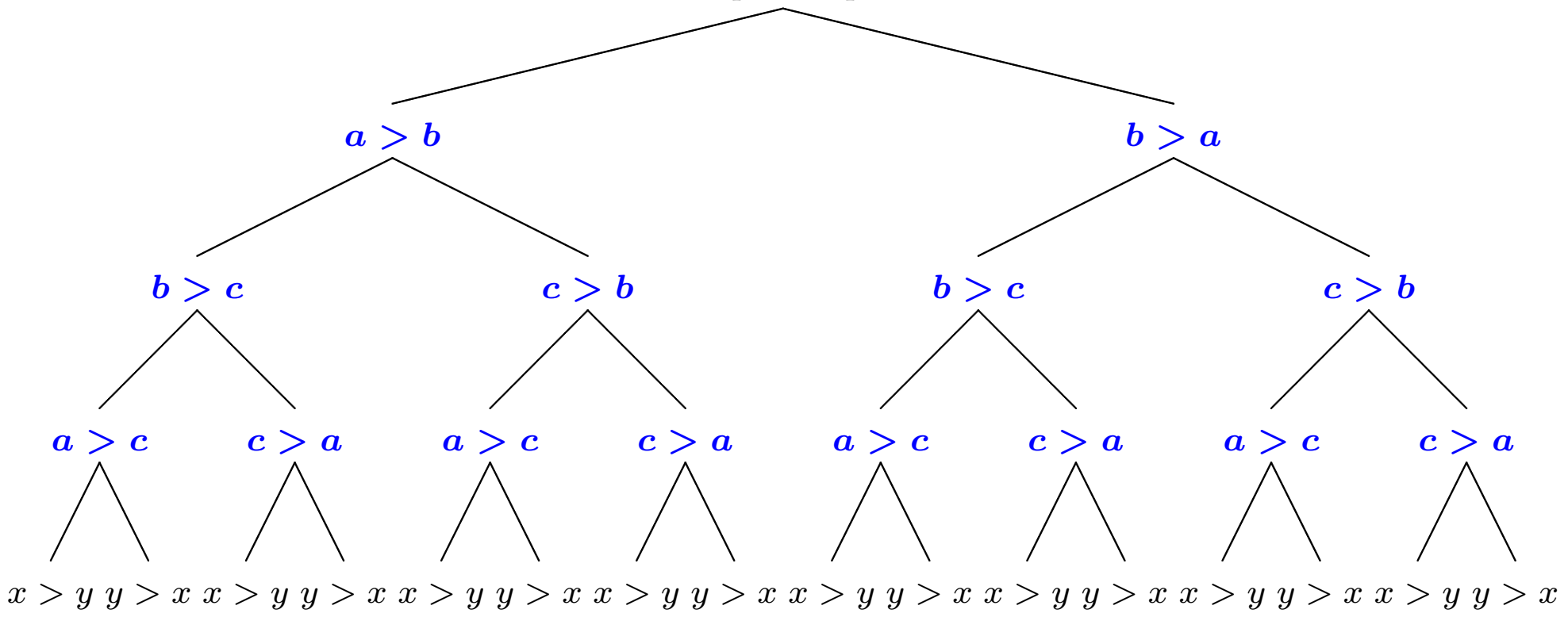

Instead of the marked subtrees, where we fix the relative orders of the pairs of $\{a, b, c\}$ successively, we create a new vertex, where we fix them at the same time. Some combinations of fixed relative orders are inconsistent (like $a > b$, $b > c$ and $c > a$). We can use this to provide only six new vertices in place of eight induced trees.

replaced node

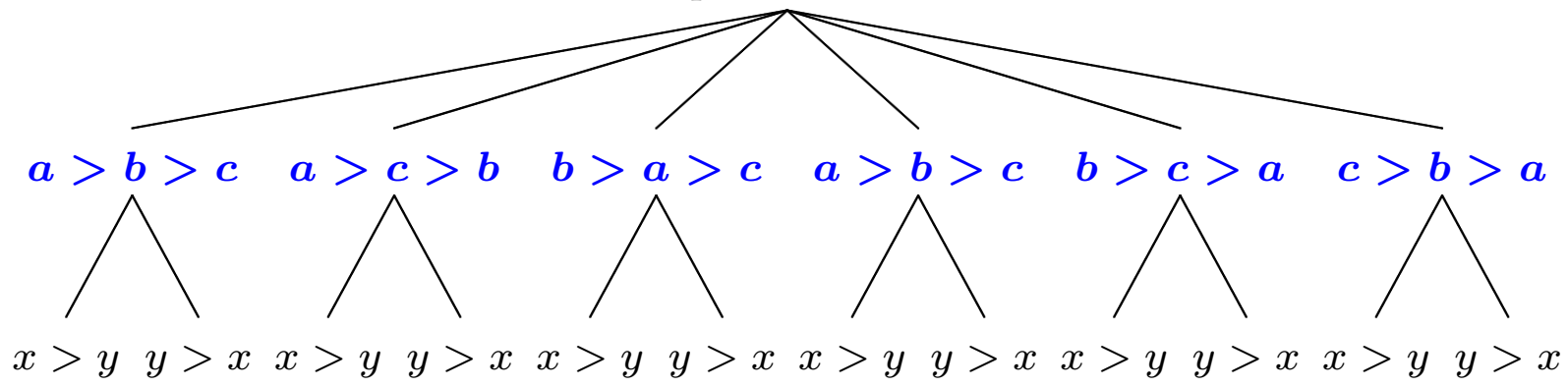

Figure 2.1.: Improving the search tree

### 2.2.1. Extending the concept of dirtiness

We start with defining "dirtiness" for a set of candidates of arbitrary size.

**Definition 4.** *Let $(V, C)$ be an election with $n$ votes and $m$ candidates. For a subset $D \subseteq C$ the* ***dirty-graph*** *of $D$ is an undirected graph with $|D|$ vertices, one for each candidate from $D$, such that there is an edge between two vertices if the corresponding candidates form a dirty pair. The subset $D$ is* ***dirty*** *if the dirty-graph of $D$ is connected. We say that $D$ is a* ***dirty $j$-set*** *if $|D| = j$ and $D$ is dirty.*

Definition 4 generalizes the concept of *dirty pairs* in Definition 2 (which is a dirty 2-set) and *dirty triples* in Section 2.1.1 (which is a dirty 3-set). We will generalize the improvement of the search tree algorithm (Figure 2.1) by branching on dirty $j$-sets with $j > 3$ instead of dirty triples.

## 2.3. The new search tree algorithm

In this subsection, we will describe the algorithm called $s$-`kconsens`.

> $s$-`kconsens`
> Program parameter: Maximal size of the analysed dirty sets $s$
> Input: An election $(V, C)$ and a positive integer $k$
> Output: A consensus list with a Kemeny score of at most $k$ or 'no'

Basically, $s$-`kconsens` works as follows. In a prebranching step it computes the set of all dirty pairs and the corresponding dirty $s$-sets. Then it branches according to the possible permutations of the candidates in the dirty $s$-sets. We only branch into cases that are possible due to Observation 4 and decrease the parameter according to Observation 4. This part of the algorithm is called branching step. If all dirty $s$-sets are handled, we fix the order of the candidates in the remaining dirty $t$-sets with $t < s$. As we will show in Section 2.3.2.1 we can use an order that minimizes the corresponding subscore. Therefore, it only takes into account permutations with consistent relation sets as discussed in Observation 4. When all relative orders are fixed, we can compute the final consensus list in polynomial time. This part of the algorithm is called postbranching step.
The following subsections are organized as follows. First, we give a more detailed description including high-level information about data structures. Second, we show the correctness and analyse the running time of $s$-`kconsens`. The theoretical analysis of the running time is restricted to the case $s = 4$.

### 2.3.1. Pseudo-code

Now, we will describe some details. The algorithm $s$-`kconsens` uses an important object $L$, that stores fixed relative orders of candidates as set of ordered pairs. We denote

```
procedure s-KCONSENS
    create new and empty L
    for each unordered pair {a, b} do
        if all votes in V rank a > b then
            L.memorize(a > b)
        end if
    end for
    return s-kconsens_rek(L)
end procedure
```

Figure 2.2.: In the initialization (prebranching step) we store the relative orders of the candidates of all non-dirty pairs. So $L$.ambiguous will return the set of dirty pairs after initialization. This initialization is correct due to Lemma 1.

this set as $L_O$. That is, $L_O := \{(x, y) \mid L \text{ stored } x > y\}$. In each storage call, $L$ determines a set of ordered pairs according to the recently fixed relative orders of candidates. That is, $L$ computes the relation set of a given permutation $L_N$ and adds it to $L_O$. Analogously to Section 2.1.2, we will denote a pair of candidates $\{a, b\}$ as *ambiguous* if $L$ does not store the relative order of $a$ and $b$. Otherwise we call it *non-ambiguous*. In a later section, we will discuss the implementation of $L$. It provides the following concrete functions:.

**L.memorize($l$)** The argument $l$ is a preference list (permutation) of candidates in $C$. It stores the relative orders of the candidates in $l$ (namely the set $L_N$). That is, $L_O \leftarrow L_O \cup L_N$. It returns 'yes' if $L_N$ and $L_O$ agree, otherwise it returns 'no'. In addition, if there is any ambiguous pair and only one order of the candidates of this pair agrees with $L_O$ it stores this relative order, too. For reference, we call this step *ambiguous-check*.

**L.ambiguous()** This function returns the set of ambiguous pairs.

**L.getList()** This function returns 'no' if there are ambiguous pairs. Otherwise it returns a preference list $p$ such that $L_O$ agrees with $p$.

**L.score()** This function returns the score implied by the stored relative orders, that is:

$$\sum_{v \in V} \sum_{\{c,d\} \subseteq C} d_v(c, d) \tag{2.1}$$

where $d_v$ is set 1 if $v$ ranks $c$ and $d$ in a different order than $L$ stored, and else $d_v$ is set 0. In other words $L$.score() computes $subscore(\bar{D})$ with $\bar{D}$ being the set of non-ambiguous pairs. Clearly, if there are no ambiguous pairs it returns the score of the uniquely determined consensus list.

The pseudo code of the algorithm is subdivided into three parts. It consists of an initialization part (Figure 2.2), a recursive part for the search tree (Figure 2.3), and some supporting functions (Figure 2.4). Now, we are able to analyse the algorithm.

```
1: function s-KCONSENS_REK(L)
2:     if L.score() > k then
3:         return 'no'
4:     end if
5:     D ← L.ambiguous()
6:     if D contains a dirty s-set D_s then
7:         for each permutation l of candidates in D_s do
8:             L_N ← L
9:             if L_N.memorize(l) = 'yes' then
10:                result ← s-kconsens_rek(L_N)
11:                if result ≠ 'no' then
12:                    return result
13:                end if
14:            end if
15:        end for
16:        return 'no'
17:    else
18:        for t = s − 1 downto 2 do
19:            for each dirty t-set D_t do
20:                best_perm(L,D_t)
21:            end for
22:        end for
23:    end if
24:    if L.score() > k then
25:        return 'no'
26:    else
27:        return L.getlist()
28:    end if
29: end function
```

Figure 2.3.: In the recursion part, we fix the relative order of the candidates by storing them in $L$. There are two cases: *Case 1 branching step* (lines 6-16): There is a dirty $s$-set. We try to store the relation set of each permutation separately. If it was possible to store the relative order of the candidates of the permutation, we call the function recursively. Otherwise (the recursive call returns 'no'), we try another permutation. If no recursive call returns 'yes' we will return 'no'. *Case 2 postbranching step* (lines 17-24): There is no dirty $s$-set. We fix the relative orders of the candidates of each dirty $s-1$-set. Thereafter we fix the relative orders of the candidates of each dirty $s-1$-set and so on. Finally we can return the consensus list if the score is not greater than $k$, else we return 'no'.

```
function PERM(D_t,i)
    Return the i'th permutation of the candidates in D_t.
end function
function BEST_PERM(L, D_t)
    scoreB ← ∞
    for i = 1 to t! do
        L_i ← L
        if L_i.memorize(perm(D_t,i)) = 'yes' then
            if L_i.score() < scoreB then
                L_B ← L_i
                scoreB ← L_i.score()
            end if
        end if
    end for
    L ← L_B
end function
```

Figure 2.4.: The support function `best_perm` stores the relation set of the permutation of $D_t$ with the best subscore for the input election, but of course it only accounts for sets that agree with $L_O$.

## 2.3.2. Analysis of the search tree algorithm

### 2.3.2.1. Correctness

We already proved in Section 2.1 that branching according to the permutations of the candidates in all dirty sets solves KEMENY SCORE. In the new algorithm, we only branch into the permutations of the candidates in all dirty $s$-sets, and compute the relative orders of the candidates in the dirty $t$-sets for $t < s$ in the search tree leaves. We have to show that it is correct to compute the best order of candidates in each dirty $t$-set without branching, that is:

**Lemma 3.** *The postbranching step of* $s$-`kconsens` *works correctly.*

*Proof.* In the postbranching step $s$-`kconsens` handles all dirty $t$-sets with $t < s$ independently, that is, it chooses the permutation with the local minimum score. We will show that for two maximal dirty $t$-sets $D_1$ and $D_2$, it must hold that for every $d_1 \in D_1$ with $d_1 \notin D_2$, $d_2 \in D_2$ with $d_2 \notin D_1$ the relative order of $d_1$ and $d_2$ is already fixed. Assume that the relative order of $d_1$ and $d_2$ is not fixed. Thus, $D_1 \cup \{d_2\}$ and $D_2 \cup \{d_1\}$ are dirty. This conflicts with the maximality of $D_1$ and $D_2$. □

In the following, we analyse the running time of $s$-`kconsens`. Therefore, we will start to find an upper bound for the search tree size. Then we will analyse the running time in the search tree nodes.

#### 2.3.2.2. Search tree size.

As mentioned before, we analyse the search tree size for $s = 4$. We can get an upper bound for the search tree size by analysing the branching number (see Section 1.3). As we already know from Section 2.1.1, the parameter decreases depending on the order of the candidates involved in dirty pairs at each search tree node. So, to get the branching vector for the search tree of $s$-`kconsens`, we do a case distinction on the number of dirty pairs in the dirty 4-set $D_4 = \{a, b, c, d\}$. Since $D_4$ is dirty, its dirty graph is connected. Each dirty pair corresponds to an edge in a connected graph with four vertices. Thus, the minimal number of dirty pairs is three and the maximal number is six.
In each case we take a look at one search tree node. The algorithm branches according to the dirty 4-set. Depending on the number of involved dirty pairs, there is a fixed number of branching cases (possible permutations for the candidates of $D_4$). We need to analyse how much the parameter decreases in each branching case. As result of Lemma 2 in Section 2.1.1, we have at least three votes. For each dirty pair, there are two cases to fix the relative order of its candidates. In the first case, the parameter will be decreased by only one. We will say, that the pair is *ordered badly* because this is the worst case for the analysis. In the second case, the parameter will be decreased by at least two. We will say, that the pair is *ordered well*. We start with discussing the cases that are easier to handle. Note, that in some cases, it would be relatively easy to find a better upper bound. We omit this since it is only reasonable to find an upper bound that is better than the upper bound in the worst case, given in Lemma 7. We will get a branching number of 1.50782 in that case.

**Lemma 4.** *If we have five dirty pairs in $D_4$, then the branching number is at most* 1.48056*.*

*Proof.* If we have five dirty pairs, one pair must have the same relative position in all votes. So, we have twelve possible permutations for the candidates of $D_4$, because in half of permutations the candidates of this pair have another relative order. In the worst case all dirty pairs are ordered badly. In this case, the parameter is decreased by $(5 \cdot 1) = 5$. Choosing one out of five pairs we have five cases, where one dirty pair is ordered well and four pairs are ordered badly. The parameter is decreased by $(1 \cdot 2) + (4 \cdot 1) = 6$ in these cases. For all other cases the parameter is decreased by at least $(2 \cdot 2) + (3 \cdot 1) = 7$, because we have at least two well ordered pairs and at most three badly ordered pairs. Thus, we have the branching vector $(5, 6, 6, 6, 6, 6, 7, 7, 7, 7, 7, 7)$. The corresponding branching number is 1.48056. □

To prove the next lemma, we introduce a new type of auxiliary graph.

**Definition 5.** *Let $(V, C)$ be an election. For a subset $D \subseteq C$ the* ***relation graph*** *of $D$ is an directed graph with $|D|$ vertices, one for each candidate from $D$, such that there is an arc from vertex $x$ to vertex $y$ if the candidate corresponding to $x$ is preferred to the candidate corresponding to $y$ in each vote.*

**Observation 5.** *The relation graph of $D$ is acyclic and contains no induced $P_3$.*

Since the first part of Observation 5 is trivial, we will only prove the second part: An induced $P_3 = (\{x, y, z\}, \{(x, y), (y, z)\})$ in a relation graph would imply that $\{x, y\}$ as well as $\{y, z\}$ have the same order $x > y$ and $y > z$ in the preference lists of all votes. Thus we have $x > z$ in all votes and also the edge $(x, z)$ in the relation graph. This conflicts with our assumption that the relation graph contains $P_3$ (and so not $(x, z)$).

**Lemma 5.** *If we have six dirty pairs in $D_4$, then the branching number is* 1.502792.

*Proof.* If all pairs of $D_4$ are dirty, we have to take into account all $4! = 24$ permutations of the candidates. Now, we will analyse how much the parameter will decrease depending on the numbers of well ordered and badly ordered pairs.

**Case 1:** Every branching possibility contains at least one well ordered pair.
Choosing one out of six pairs there are at most six cases with only one well ordered pair and five pairs are ordered badly. The parameter is decreased by $(1 \cdot 2) + (5 \cdot 1) = 7$ in these cases. Choosing two out of six pairs, we have 15 other cases with two well ordered and four badly ordered pairs. The parameter will decrease by $(2 \cdot 2) + (4 \cdot 1) = 8$. In the remaining three cases, the parameter is decreased by at least $(3 \cdot 2) + (3 \cdot 1) = 9$. This causes a branching vector of $(7, 7, 7, 7, 7, 7, 8, 8, 8, 8, 8, 8, 8, 8, 8, 8, 8, 8, 8, 8, 8, 9, 9, 9)$. So, the branching number is 1.502792.

**Case 2:** There is a branching possibility that contains only badly ordered pairs.
Trivially, choosing one out of six pairs there are at most six cases with only one well ordered pair. We will now show that there are at most four such cases. We get all possible cases by assuming that all pairs are ordered badly and flipping the order of one single pair. We will show, that this is only possible for four of the six pairs: We already know that ordering all pairs badly will cause no cycle in the relations graphs of the subsets of $D_4$: $\{a, b, c\}$, $\{a, d, c\}$ and $\{b, c, d\}$.
*Claim:* Flipping the order of (at least) two of the six pairs causes a cycle in the relation graph of $D_4$.
*Proof:* For each of the three sets above flipping the order of one pair causes a cycle in the relations-representing graph[1]. We denote this pair as *cycle pair*.

**Case 2.a** For $\{a, b, c\}$ and $\{a, d, c\}$ the cycle pair is the same. Thus, it must hold $s_1 = \{a, c\}$. Then, for $\{b, c, d\}$ we have another pair $s_2 \in \{\{b, c\}, \{c, d\}, \{b, d\}\}$ with $s_2 \neq s_1$.

**Case 2.b** For $\{a, b, c\}$ and $\{a, d, c\}$ the cycle pair is not the same. Then, we already have two different pairs.

Thus, flipping the order of at most four pairs will cause a consistent relation set. Hence, we actually have at most four cases, where only one pair is ordered well. Analogously to Case 1, we have another 15 cases with two well ordered pairs. Thus, we have the branching vector $(6, 7, 7, 7, 7, 8, 8, 8, 8, 8, 8, 8, 8, 8, 8, 8, 8, 8, 8, 8, 9, 9, 9, 9)$. The corresponding branching number is 1.502177.

[1]Assume that $x > y > z$. This implies $x > y$, $y > z$, and $x > z$. We can flip $y > z$ to $z > y$ so that $x > z > y$ and we can flip $x > y$ to $y > x$ so that $y > x > z$. Flipping only $x > z$ to $z > x$ would mean a cycle in the relations graph of $\{x, y, z\}$: $z \to x \to y \to z$. This contradicts Observation 5.

| | | |
|---|---|---|
| 1. $a > b > c > d$ | 9. $c > a > d > b$ | 17. $b > c > d > a$ |
| 2. $a > b > d > c$ | 10. $d > a > c > b$ | 18. $b > d > c > a$ |
| 3. $a > c > b > d$ | 11. $c > d > a > b$ | 19. $c > b > a > d$ |
| 4. $a > d > b > c$ | 12. $d > c > a > b$ | 20. $d > b > a > c$ |
| 5. $a > c > d > b$ | 13. $b > a > c > d$ | 21. $c > b > d > a$ |
| 6. $a > d > c > b$ | 14. $b > a > d > c$ | 22. $d > b > c > a$ |
| 7. $c > a > b > d$ | 15. $b > c > a > d$ | 23. $c > d > b > a$ |
| 8. $d > a > b > c$ | 16. $b > d > a > c$ | 24. $d > c > b > a$ |

Figure 2.5.: Table of permutations

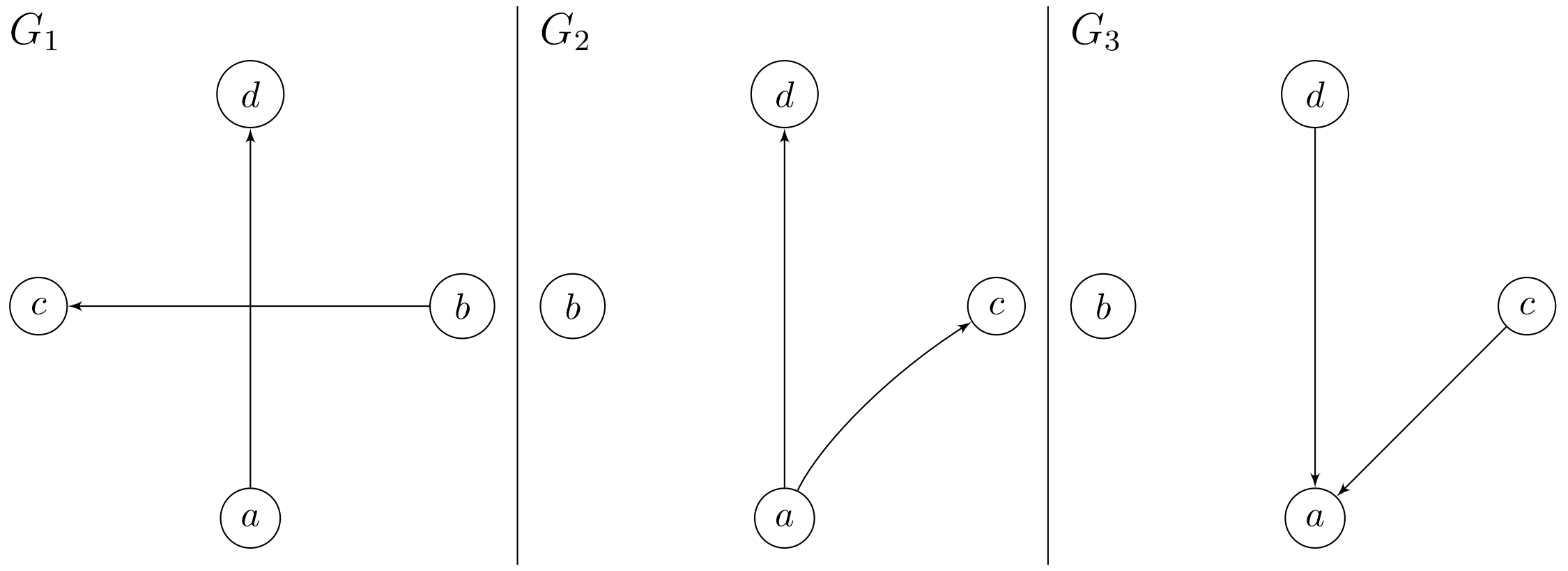

Figure 2.6.: Relation graphs of $D_4$ if two pairs have fixed order.

□

**Lemma 6.** *If we have four dirty pairs in $D_4$, then the branching number is* 1.496327*.*

*Proof.* Two pairs must have the same preference list in all votes, because only four of six pairs are dirty. Now we will look at the relation graphs of $D_4$ (Figure 2.6). According to Observation 5, there is no induced $P_3$. Thus, there are up to isomorphism only 3 possible relation graphs of $D_4$. Either the pairs are independent (see $G_1$) or they share one candidate (see $G_2$ and $G_3$).
Now we will analyse each possible relation graph of $D_4$.

**$G_1$:** The relative orders $a > d$ and $b > c$ are fixed.
For this case the permutations $3, 5, \ldots, 12, 16, \ldots, 24$ (see Figure 2.5) are not possible. Only six permutations are left over. A simple calculation analogous to Lemma 4 gives the branching vector $(4, 5, 5, 5, 5, 6)$. Thus, the branching number is 1.437259.

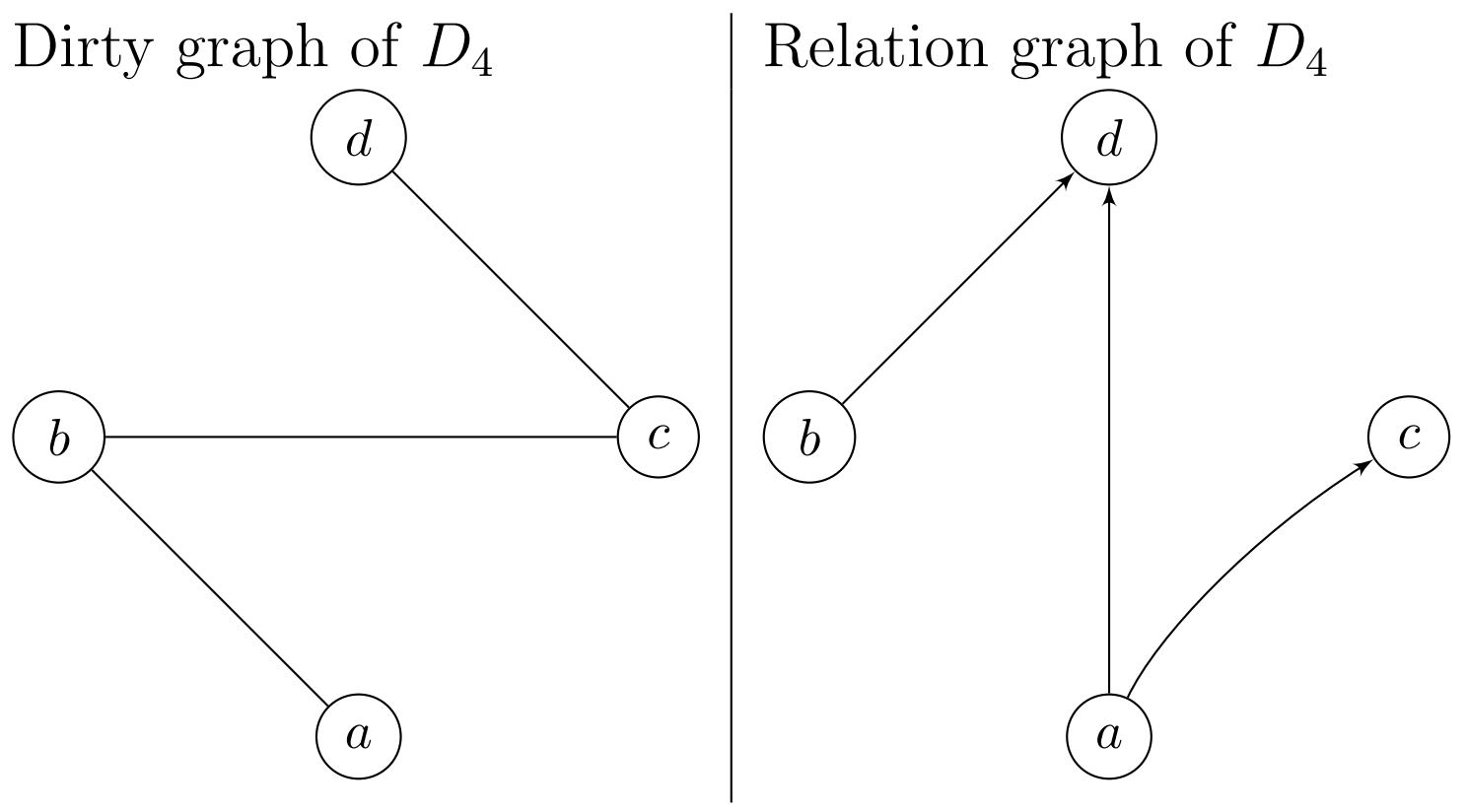

Figure 2.7.: The dirty graph of $D_4$ and the relation graph of $D_4$ for three pairs with fixed order.

**$G_2$:** The relative orders $a > d$ and $a > c$ are fixed.
For this case the permutations $7, \dots, 12, 15, \dots, 24$ (see Figure 2.5) are not possible.

**$G_3$:** The relative orders $d > a$ and $c > a$ are fixed.
For this case the permutations $1, \dots, 10, 13, \dots, 16, 19, 20$ (see Figure 2.5) are not possible.

In both graphs $G_2$ and $G_3$ only eight permutations are left over. Analogous to $G_1$, we get the branching vector $(4, 5, 5, 5, 5, 6, 6, 6)$. The branching number is 1.496327 □

For the proof of the next lemma, we introduce another type of auxiliary graph.

**Definition 6.** *For a subset $D \subseteq C$ the* ***election multigraph*** *of $D$ is a directed multigraph with $|D|$ vertices, one vertex for each candidate from $D$, such that for each vote there is an arc from vertex $x$ to vertex $y$ if the candidate corresponding to $x$ is preferred to the candidate corresponding to $y$.*

**Lemma 7.** *If there are three dirty pairs in $D_4$, then the branching number is* 1.50782*.*

*Proof.* In this case, three pairs are fixed and three pairs are dirty. So, up to isomorphism, there is only one dirty graph of $D_4$ (see Figure 2.7), since the graph has to be connected. Further, all three non-dirty pairs have the same order in all votes and induced $P_3$s are not allowed due to Observation 5. More precisely, if $a$ is preferred to $c$ then $a$ must be preferred to $d$, because $\{d, c\}$ is dirty. Furthermore, $b$ must be preferred to $d$, because $\{b, a\}$ is dirty. Assuming $c > a$ leads to an isomorph graph. Thus, up to isomorphism, there is only one relation graph of $D_4$ (see Figure 2.7).
According to this relation graph of $D_4$, all votes rank $a > c$, $a > d$ and $b > d$. If we have these relations fixed, only the following five permutations are left over.

$$P_1 : a > b > c > d$$
$$P_2 : a > b > d > c$$
$$P_3 : a > c > b > d$$
$$P_4 : b > a > c > d$$
$$P_5 : b > a > d > c$$

Now we want to analyse the branching vector. To this end, we show that the branching vector is at least as good as $(3, 4, 4, 4, 5)$ for all inputs. We will use again the fact that there are more than two votes.
We do a case distinction on the election multigraphs of $D_4$ to get the worst case branching number. There are five possible election multigraphs of $D_4$. To see this, take a look at Figure 2.8. We can simply count for each permutation for each pair how many votes rank them in a different order. The following table shows how much the parameter decreases for each permutation ($P_1$-$P_5$) for each election multigraph of $D_4$:

| | $P_1$ | $P_2$ | $P_3$ | $P_4$ | $P_5$ |
|---|---|---|---|---|---|
| $G_1$ | 3 | 4 | 4 | 4 | 5 |
| $G_2$ | 4 | 3 | 4 | 5 | 4 |
| $G_3$ | 4 | 5 | 3 | 5 | 6 |
| $G_5$ | 4 | 5 | 5 | 3 | 4 |
| $G_6$ | 5 | 4 | 6 | 4 | 3 |

The worst case branching vector is $(3, 4, 4, 4, 5)$. Thus, the branching number is at most 1.50782. □

**Lemma 8.** *The search tree size is $O(1.50782^k)$ with $k$ being the Kemeny score.*

*Proof.* Due to the Lemmas 4-7 the worst case yields the branching number 1.50782. We get a search tree of size $O(1.50782^k)$. □

This is an improvement from $O(1.53^k)$ to $O(1.5078^k)$ for the worst-case search tree size for dirty 4-sets. There is hope that branching on dirty $s$-sets with $s > 4$ will improve the worst-case running time further. We will test the algorithm in practice in the next chapter. Notably, the implementation is much easier than its theoretical analysis. There is no big overhead for arbitrary $s$ (maximum size of analysed dirty sets) for the polynomial factor of the running time as we discuss next.

#### 2.3.2.3. Running time

At this point, we will analyse the running times of the prebranching, branching (polynomial part) and postbranching steps of the algorithm to get the overall running time. In the prebranching step, the algorithm enumerates the dirty pairs and precomputes the subscore of each pair. There are $m \cdot (m - 1)$ ordered pairs and $n$ votes. Thus, this is done in $O(m^2 \cdot n)$ time. To improve the running time of the branching-step, $s$-`kconsens` precomputes the set of dirty sets in this step. It finds the dirty sets by iterating over all dirty pairs and builds up the sets step by step. Because it only has to mark each (unordered) dirty pair once, this takes $O(m^2)$ time. In the branching-step we have to

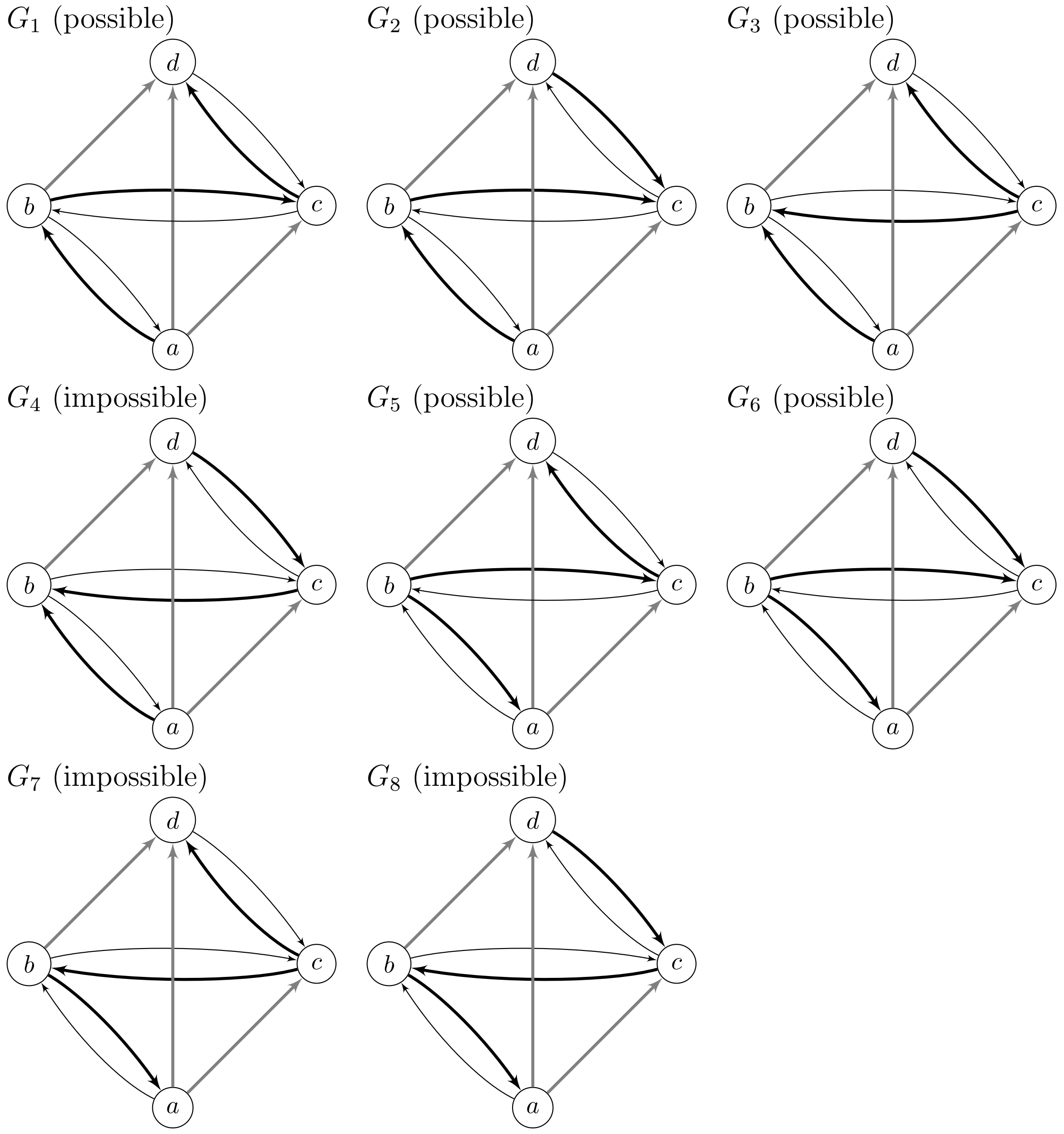

Figure 2.8.: The simplified election multigraph of $D_4$ for three pairs with fixed order. Since we use the fact that we have at least three votes, we will draw a thin arrow from $d_i$ to $d_j$ if there is only one vote in $V$ with $d_i > d_j$. A fat arrow from $d_i$ to $d_j$ denotes that there are at least two votes with $d_i > d_j$ and grey arrows denote that $d_i > d_j$ in all votes. To see which graph is possible, take a look at Figure 2.9.

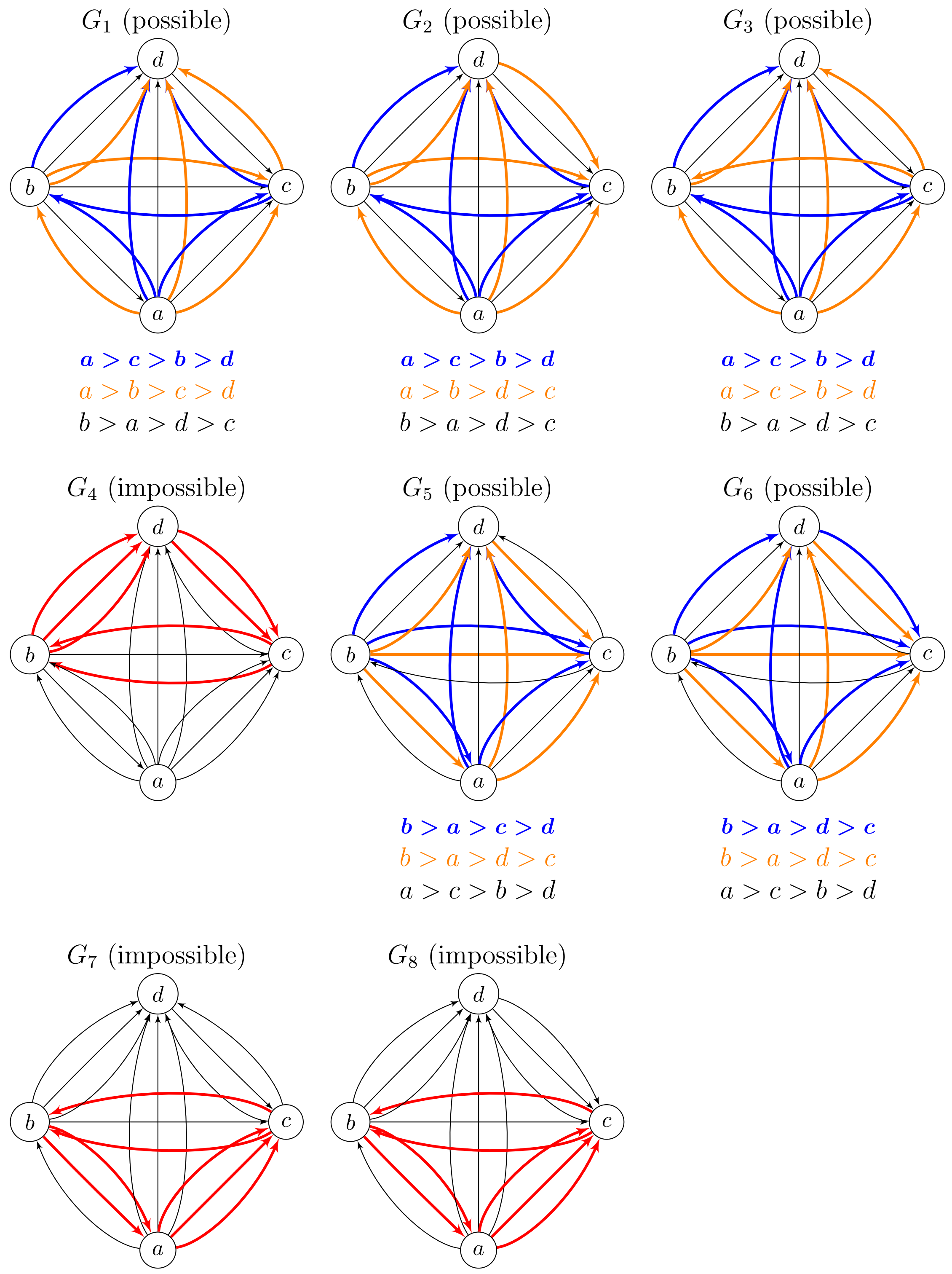

Figure 2.9.: The election multigraphs of $D_4$ for three pairs with fixed order. For each possible election multigraph it must be possible to assign one arc between each vertex pair to each vote. In $G_4$, $G_7$, and $G_8$ it is not possible to assign the arcs to the votes without assigning a cycle.

analyse the polynomial running time in the search tree nodes. At each search tree node, $s$-`kconsens` fixes the relative orders of at most $s \cdot (s-1)/2$ dirty pairs for one permutation. This is done in constant time for one pair. Thus, the running time at each search three node is constant, for fixed $s$. In the postbranching step, $s$-`kconsens` fixes the relative orders of the dirty pair involved in dirty $t$-sets for $t < s$. Therefore it checks less than $O(m^2)$ possible permutations (isolated). Building up a consensus list from the fixed relations is also done in $O(m \cdot (m-1))$ time. Summarizing, this leads together with Lemma 8 to the following theorem:

**Theorem 1.** *The algorithm $s$-`kconsens` computes* KEMENY SCORE *in* $O(1.5079^k + m^2 \cdot n)$.

## 2.4. Data reduction

In this section, we want to analyse some additional improvements. Therefore, we will examine data reduction rules, another discard criterion for the search tree, and a polynomial-time solvable special case. Betzler et al. [BGKN09] used in very recent studies another characterization of dirtiness. We will use it and rename this concept as *majority-dirtiness*. Let $(V, C)$ be an election as discussed before. An unordered pair of candidates $\{a, b\} \subseteq C$ with neither $a > b$ nor $a < b$ in more than 2/3 of the votes is called *majority-dirty pair* and $a$ and $b$ are called *majority-dirty* candidates. All other pairs of candidates are called *majority-non-dirty pairs* and candidates that are not involved in any majority-dirty pair are called *majority-non-dirty candidates*. Let $D_M$ denote the set of majority-dirty candidates and $n_M$ denote the number of majority-dirty pairs in $(V, C)$. For two candidates $a, b$, we write $a >_{2/3} b$ if $a > b$ in more than 2/3 of the votes. Further, we say that $a$ and $b$ are *ordered according to the* 2/3 *majority* in a preference list $l$, if $a >_{2/3} b$ and $a > b$ in $l$. Betzler et al. [BGKN09] showed the following theorem:

**Theorem 2.** *[BGKN09]* KEMENY SCORE *without majority-dirty pairs is solvable in polynomial time.*

If an instance has no majority-dirty pair, then all candidate pairs in every Kemeny consensus are ordered according to 2/3-majority. We can easily find the corresponding consensus list in polynomial time. This means, we have another polynomial-time solvable special case. We can identify this case in $O(m^2 \cdot n)$. Therefore we check for each candidate pair if the relative order of the candidate is the same in more than 2/3 of the votes. It would be interesting to use this result in the search tree, too. A promising possibility would be that a search tree algorithm fixes the relative order of the candidates of each majority-dirty pair. All majority-non-dirty pairs would be ordered according the 2/3-majority. Unfortunately, it is not clear that there should be any preference list that agrees with the resulting set of fixed ordered pairs. However, in the following lemma, we note another interesting fact, leading to an idea for a data reduction rule. To this end, we need the term *distance* between two majority-non-dirty candidates. For a majority-non-dirty pair $\{c, c'\}$ we define $\text{dist}(c, c') := |\{b \in C : b$ is majority-non-dirty and

$v_1$: $y > a > b > c > d > x$
$v_2$: $y > a > b > c > d > x$
$v_3$: $c > d > x > y > a > b$
$v_4$: $a > d > x > y > b > c$
$v_5$: $a > b > x > y > c > d$
$v_6$: $b > c > x > y > a > d$
$v_7$: $a > b > c > d > x > y$

Figure 2.10.: Votes of the election in Example 6.

$c >_{2/3} b >_{2/3} c'\}|$ if $c >_{2/3} c'$ and $\text{dist}(c, c') := |\{b \in C : b$ is majority-non-dirty and $c' >_{2/3} b >_{2/3} c\}|$ if $c' >_{2/3} c$.

**Lemma 9.** *[BGKN09] Let $(V, C)$ be an election and $D_M$ its set of majority-dirty candidates. If for a majority-non-dirty candidate $c$ it holds that $dist(c, c_d) > 2n_M$ for all $c_d \in D_M$, then in every Kemeny consensus $c$ is ordered according to the $2/3$-majority with respect to all candidates from $C$.*

However, note that the argumentation of Lemma 9 cannot obviously be carried over to the case that the order of some candidate pairs is already fixed. (See constrained ranking [vZW07, vZHJW07]) Thus, it is not possible to apply the corresponding data reduction rule at the search tree nodes. This can also be seen by the following example.

**Example 6**
We have an election $(V, C)$ with $V = \{v_1, v_2, \ldots, v_7\}$ and $C = \{a, b, c, d, x, y\}$. The votes are given in Figure 2.10. The candidates $a, b, c, d$ and $y$ are majority-dirty. The majority-non-dirty pairs are illustrated in Figure 2.11. We see that only the candidate $a$ is majority-non-dirty. Due to transitivity, there is only one preference list, that agrees with the order of the majority-non-dirty pairs. $(a > b > c > d > x > y)$ This preference list has a score of 34. In contrast, the Kemeny score of that election is 33, for example with $y > a > b > c > d > x$. This means, that the majority-non-dirty candidate $x$ is not ordered according the $2/3$-majority with respect to all candidates from $C$ (only with respect to the candidates $a, b, c$ and $d$).

We can see in Example 6 that there are instances, where each optimal consensus has at least one majority-non-dirty pair that is not ordered according its $2/3$ majority. This example also confirms the existence of constrains in Lemma. 9. However, there is another approach that we can use to improve our search tree algorithm:

**Lemma 10.** *[BGKN09] For an election containing $n_M$ majority-dirty pairs, in every optimal Kemeny consensus at most $n_M$ majority-non-dirty pairs are not ordered according to their $2/3$-majorities.*

We can use this as another criterion to discard some possibilities in the branching early. More precisely we can stop branching by trying all possibilities with at most $n_M$ majority-non-dirty pairs that are not ordered according to their $2/3$-majorities. In the next chapter, we will see that there is a heuristic that also guarantees that we will never fix more than $n_M$ majority-non-dirty pairs not ordered according to their 2/3-majority. This means, we do not have to implement an additional termination condition for Lemma 10. Lemma 9 says that in every Kemeny consensus all pairs, containing a majority-non-dirty candidate, are ordered according to the 2/3-majority under certain conditions. Another possibility for restricting conditions is to check if there are Condorcet winners or Condorcet loosers and remove them from the instance. This leads to the following reduction rule.

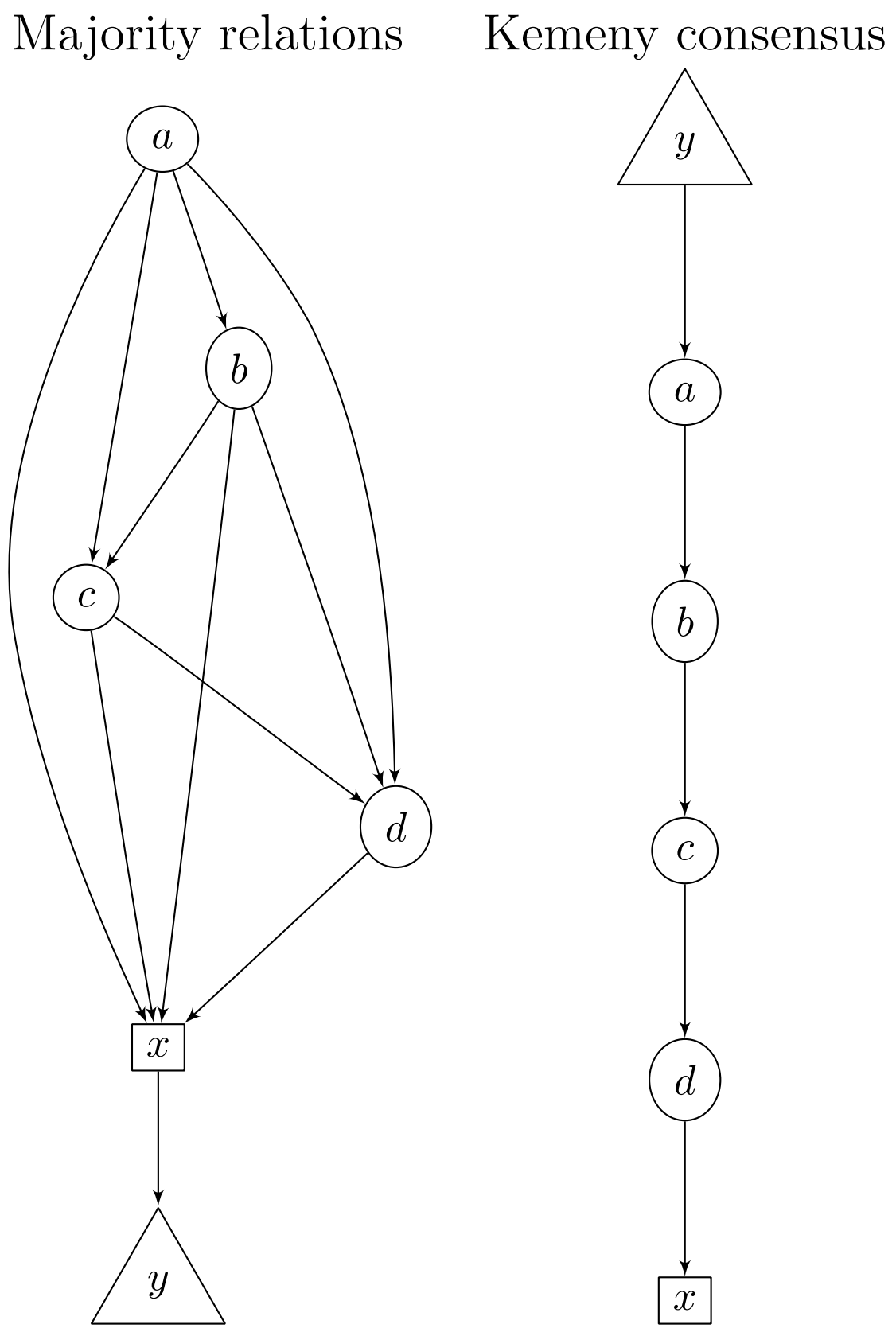

Figure 2.11.: We have a directed graph with 7 vertices, one for each candidate from Example 6, such that there is an arc from vertex $x$ to vertex $y$ if the candidate corresponding to $x$ is preferred to the candidate corresponding to $y$ in 2/3 of the votes.

**Reduction rule "Condorcet winner/looser"**

Let $c$ be a non-dirty candidate. If $c$ is most preferred (least preferred) in more than half of the votes, then delete $c$ and decrease the Kemeny score by the subscore of the set of candidate pairs containing $c$.

Note that using this reduction rule also covers the special case of Theorem 2. After exhaustively applying the reduction rule on an instance without majority-dirty pairs, the number of candidates is zero. Thus, we have solved that instance without branching. The reduction rule works correctly, because Kemeny's voting scheme satisfies the Condorcet criterion. It is trivial to see, that it takes polynomial time to apply the data reduction rule: It's take $O(m^2 \cdot n)$ to get all non-dirty pairs. Thereafter, it takes $O(m)$ for each candidate, to check if he is most preferred or least preferred in half of the votes. In further studies, it should be possible to extend this reduction rule by searching for a "set of winners" ("set of loosers").

# 3. Implementation and experiments

In the preceding chapter, we saw that the new search tree algorithm has a better theoretical running time than the less refined variants. Now, we will show that this improvement also pay off in practice. We implemented and tested algorithms for two parameterizations. A dynamic programming algorithm for the parameter $m :=$"number of candidates" was developed in [BFG+08] and runs in $O(2^m \cdot m^2 \cdot n)$ time. We call it `kconsens_cands`. For the sake of completeness it is described in the next paragraph. For comparison we also implemented the less refined search tree algorithms. The trivial search tree algorithm is called `kconsens_pairs` and the improved search tree algorithm of [BFG+08] is called `kconsens_triples`. They are described in Section 2.2. Finally, we implemented the algorithm $s$-`kconsens` for arbitrary $s$. All these implementations are tested on randomly generated and real-world data as discussed in the following. We will see that they are fast enough to use them in practice for the considered test data. Thus, we have a fast method of computing an optimal consensus list for some realistic instances. Another part of the experiments is the analysis of some properties of the instances. Besides obvious properties like "number of votes" or "number of candidates" we take also into account the "number of dirty pairs" and the "number of majority-non-dirty pairs". Clearly, they influence the running time of the search tree algorithms and the applicability of the data reduction rule "Condorcet winner/looser". We also determine an upper and a lower bound for the Kemeny score. Further, we investigate the values of the "maximum range of two candidates", defined as $\max_{v,w \in V; c,d \in C} |\operatorname{pos}_v(c) - \operatorname{pos}_w(d)|$, and the "average kt-distance between the input votes" because there exist parameterized algorithms that use these parameters [BFG+09]. Another property can be considered: the number of candidates that could be removed after exhaustively applying of the data reduction rule in the preprocessing. An overview of all properties can be found in the beginning of the appendix.

**Parameterized algorithm with the parameter "number of candidates"** Simply trying all permutations of candidates already leads to fixed-parameter tractability with respect to this parameter [BFG+08]. Unfortunately, its corresponding running time $O(m! \cdot mn \cdot logm)$ is not practicable. We briefly describe a dynamic programming algorithm, which outputs an optimal Kemeny consensus for a given election $(V, C)$. For each subset of the candidates set compute the Kemeny score restricted to this subset, that is, the subscore of the set of candidate pairs from this subset. The recurrence for a given subset $C'$ is to consider every subset $C'' \subseteq C'$ where $C''$ is obtained by deleting a single candidate $c$ from $C'$. Let $l''$ be a Kemeny consensus for the election restricted to $C''$. Compute the score of the permutation $l'$ of $C'$ obtained from $l''$ by putting $c$ in the first position. The algorithm takes the permutation with the minimum score over of all $l'$ obtained from subsets of $C'$ as consensus. Its running time is in $O(2^m \cdot m^2 \cdot n)$.

## 3.1. Implementation

In this chapter, we want to show the effectiveness of the algorithms in practice. Therefore, a high-performance implementation is necessary. We decided to use C++ as programming language. First of all, it is very popular and many programmers are able to read C++. Furthermore, there are many high-performance libraries available for complex and mathematical computation. In our implementation we use several libraries of the popular "boost" library package [boo09]. Our project has got 14 classes and 3601 lines of code (without comments) overall. Besides an intelligent memory management and a high-performance data-structure for the subsets the implementation of `kconsens_cands` is implemented straight forward to the description. The implementations of `kconsens_pairs` and `kconsens_triples` are close to their original description in [BFG+08] and Section 2.2. For $s$-`kconsens` we have the pseudo-code in Section 2.3.1, which describes the algorithmic details. There is some additional information about the implementation of $s$-`kconsens` in the next paragraph.

**Details of the implementation of $s$-`kconsens`** Here, we describe some details of the implementation of $s$-`kconsens`. Some considerations turned out to be helpful to improve the running time in practice, although their implementation influences the theoretical worst case running time by a polynomial factor to the search tree size. We again briefly discuss the running time.
Unfortunately, the recursive description in the pseudo code is a little bit improper for high performance (in C++). Thus, we transform the recursive part of the algorithm into an iterative form. The results for often called queries like "subscores of a permutation" or "is this pair dirty" are precomputed and stored in a hash map or arrays with an intelligent index system such that requests take constant time. The stack of the recursive calls is simulated by an array of data structures. For faster access and memory savings all candidates (names) will be mapped to integers in a preprocessing step. After the computation of the Kemeny score and consensus list, the original candidate names will be restored. Beside precomputing the dirty sets, in the implementation $s$-`kconsens` also precomputes the subscores of each permutation of each dirty set. There are $s!$ permutations for at most $m \cdot (m-1)/2$ dirty sets. Thus, this is done in $O(s! \cdot (m \cdot (m-1)/2))$. As discussed in Observations 3 and 4, we have two criteria to discard a branching. Therefore $L$ was implemented as a data structure that manages for each candidate $x$ two hash sets, one for candidates that are preferred to $x$ and one for the candidates, where $x$ is preferred to them. The fixing of the relative order of the candidates of one candidate pair with $L$.memorize() now takes $O(m)$ instead of constant time. (There are $4m$ hash-sets and each hash-set has to be updated at most once.) In return, $s$-`kconsens` checks the consistence in constant time at each search tree nodes. This improved the performance in practice. Since it has precomputed the subscores of the permutations, the algorithm sorts them by subscore and tests the permutations with small subscore at first. Thus, if we discard a branching due to Observation 3, then we can also skip the permutations with a greater subscore. Another point is that we call $s$-`kconsens` with $k$ being the minimal sum of pairwise subscores as a lower bound. If it returns 'no' we increase $k$ by one and call $s$-`kconsens` again. At this point we can guarantee, that we do not need to use Lemma 10 to discard the branching: If there are $n_o$ pairs fixed, not ordered according to their 2/3-majority, then there is no consensus with less than $n_o$ pairs not ordered according their 2/3 majority with less score. We already know that there is no solution with less score, because we tested theirs existence in the last call of $s$-`kconsens` and we started with a lower bound for $k$, where it is only possible that all pairs are ordered according to their 2/3-majority.

## 3.2. Data and experiments

We use several different sources to get test instances for the algorithms. The first type are randomly generated instances, which are very useful to produce performance diagrams that show the dependency of the running time on miscellaneous attributes. The second type are results of sports competitions as also discussed in the introduction. Besides Formula One, we also used several cross-skiing and biathlon competitions. In this context, apart from the running time and several attributes of the instances also the comparison of the consensus list with the results of the original point scoring system may be interesting. Last, but not least, we consider one of the most famous applications of modern rank aggregation, that is, meta search engines. The result of different search request form the votes of our rank aggregation problem. We will generate several instances, analyse their properties and test the performance of our algorithms.

### 3.2.1. Randomly generated instances

Generating random data for testing algorithms is very popular and dangerous at the same time. The significance of the tests depends on the probability space, the parameter values and the way we are using the random data. There are known cases where algorithms are provable very efficient on randomly generated instances, but do not perform well on general instances. An example is the HAMILTON CYCLE problem, which is NP-hard in general, but easy on a special class of random graphs. It is described in [MU05, Section 5.6.2]. Nevertheless, we need several series of parameter values.

The data generation works as follows: We start with generating one reference vote. Then we use this reference vote to generate all other votes by swaping some candidate pairs. To this end, we define some parameters:

1. the number of candidates ($m$)
2. the number of votes ($n$)
3. the expected number of swaps per vote ($w$)

4. the maximum distance of the swap candidates with respect to the reference vote ($d$)

Note that Conitzer et al. [Con06] also used some random data to test their algorithms. They generated a total order representing a consensus ordering. Then, they generated the voting preferences, where each one of the voters agrees with the consensus ordering regarding the ranking of every candidate pair with some consensus probability. Using the same probability for each candidate pair to be dirty, like they did, seems to generate "isolated" dirty pairs unusually often. This would be an unrealistic advantage for our search tree algorithms. Hence, we used our own way of generating the data as described above.

**Properties** For the first test series we generated instances with a growing number of candidates. Since we want to investigate the relation between the number of candidates and the running time, we fixed the rate of dirty candidate pairs, so that approximately half of all candidate pairs are dirty. For the second test series we generated instances with a constant number of candidates and a growing number of dirty pairs. This was done by varying the number of votes and the values of the 3rd and the 4th parameter in the generating process. The parameter values, used to create the test instances, in the appendix in Section A.2.1. The decision for 14 candidates in the second test series is to limit the overall running time of the test series and provide sufficiently large range of possible values for the number of dirty pairs. Other values lead to similar results.

**Results** For both series the algorithms `kconsens_cands`, `kconsens_pairs`, `kconsens_triples`, and `4-kconsens` were tested. One can except for the first test series that `kconsens_cands` will be more efficient than the search tree algorithms: It has a running time of $O(2^m \cdot m^2 \cdot n)$ while the best search tree algorithm takes $O(1.53^k + m^2 \cdot n)$ time. A lower bound for $k$ is the number of dirty pairs. Unfortunately, the number of candidate pairs is proportional to $m^2$. We can see the results from test series one in Figure 3.1. They come up to one's expectations: `kconsens_cands` is the most efficient. The algorithm `kconsens_pairs` is the least efficient while an improvement of the running time when branching over dirty 4-sets instead of dirty triples is noticeable. The results for the second test series are illustrated in Figure 3.2. Here, we can see that the search tree algorithms are significantly more efficient for these instances. Also using greater dirty sets for branching improved the running time considerable.

In summary the tests show that both parameterizations are practicable for specific types of instances. While one should use `kconsens_cands` for all instance with only a few (up to thirty) candidates, the search tree algorithms seems to be very efficient for instances with a low number of dirty pairs. In the following, we will see that this also applies to real world aggregation data.

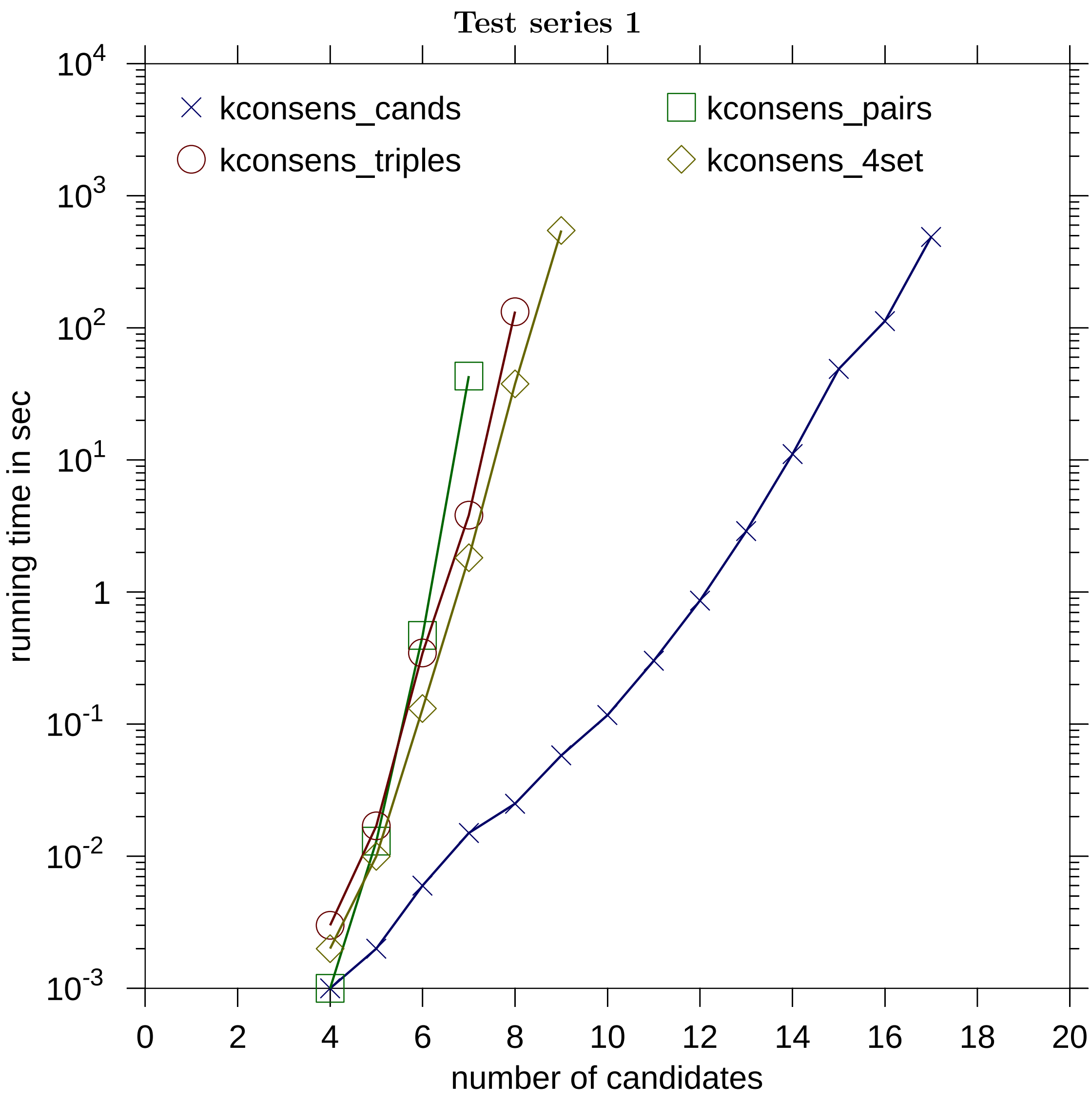

Figure 3.1.: Randomly generated data: Running time against the number of candidates. For each number of candidates ten instances where generated and tested. In all instances about 50% of the candidate pairs are dirty. We computed the average values to get more significant results. A single test run was canceled if it took more than one hour. The test series for each algorithm was canceled if the total running time for the instances of with the same number of candidates was greater than two hours.

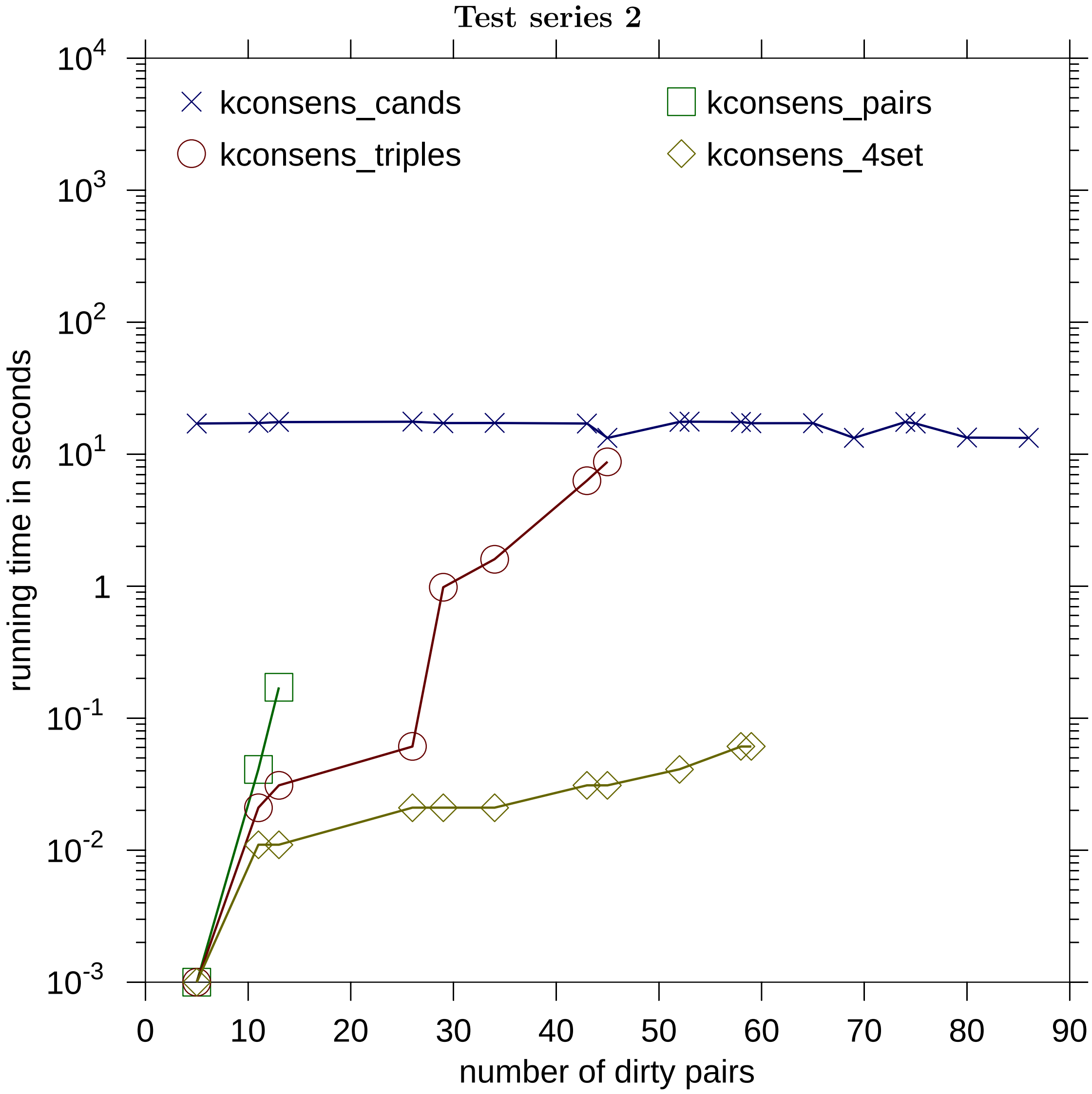

Figure 3.2.: Randomly generated data: Running time against number of dirty pairs. Here a test run was canceled if it took more than 30 minutes. All instances have 14 candidates. The number of votes and swaps as well as the swap range can be found in the Appendix.

## 3.2.2. Sports competitions

### 3.2.2.1. Formula One

As discussed in the introduction, sports competitions naturally provide ranking data. One famous sports is motor sports, especially Formula One. We generated ranking data from the Formula One seasons of the years 1961 till 2008, with one candidate for each driver and one vote for each race. The preference lists comply with the order of crossing the finish line. All drivers who fail are ordered behind the others (and their order complies with the elimination order). For the sake of simplicity, the algorithms were designed to deal only with complete preference lists without ties. Therefore, we removed the drivers who did not attend all races. In most of the seasons only about two or three candidates were removes.

**Properties** We analysed the properties, discussed in the introduction of this chapter, for the Formula One instances. Unfortunately, analysis of the instances showed that 90-100% of the candidate pairs are dirty. Moreover, the maximum range of the candidates is circa 95% of the number of candidates, which is the maximum possible value. This seems to be hard for the algorithms `kconsens_pairs`, `kconsens_triples`, and $s$-`kconsens` as we will see in the results. The data reduction rule "Condorcet winner/looser" could remove one candidate from the instances, created from the Formula One seasons 1963, 1980-81, 1992, 2001-02, and two candidate for the instance, created from the season 2004. A complete table of properties can be found in the Appendix A.1.1.

**Results** We tested the algorithms `kconsens_cands`, `kconsens_pairs`, `kconsens_triples`, and $s$-`kconsens` for $s \in \{4, \ldots, 6\}$ for the generated elections. We are not able to compute the Kemeny Score with the search tree algorithms for many instances in less than three hours. However, at least with `kconsens_cands` we are able to compute the optimal consensus list for almost all Formula One seasons in a few hours. So, the FIA could use the Kemeny voting system in prospective seasons. We compare the Kemeny consensus of the election, created from the result of a season, with the preference list, computed by the point scoring system of the FIA. For instance, we compare all drivers who attended all races in the season of 2006 in Table 3.1. As we can see here (and also if we compare the preference lists for other seasons), the preference lists are similar, especially for the drivers that get points in most of the races. Although the world champion would not change in most of the seasons, the Kemeny consensus ranks some of the drivers differently in each season. For mathematical purposes, the Kemeny consensus is more balanced in the sense that it weights each pairwise comparison equally. Otherwise, it is of course the decision of the FIA to weight the winner of a race disproportionately high. We will see all results of the Formula One generated instances in Section A.1.1 in the appendix.

| Points | FIA ranking | Kemeny consensus |
| --- | --- | --- |
| 134 | Fernando Alonso | Fernando Alonso |
| 121 | Michael Schumacher | Michael Schumacher |
| 80 | Felipe Massa | Felipe Massa |
| 72 | Giancarlo Fisichella | Kimi Räikkonen |
| 65 | Kimi Räikkonen | Giancarlo Fisichella |
| 56 | Jenson Button | Jenson Button |
| 30 | Rubens Barrichello | Rubens Barrichello |
| 23 | Nick Heidfeld | Nick Heidfeld |
| 20 | Ralf Schumacher | Jarno Trulli |
| 15 | Jarno Trulli | David Coulthard |
| 14 | David Coulthard | Ralf Schumacher |
| 7 | Mark Webber | Vitantonio Liuzzi |
| 4 | Nico Rosberg | Scott Speed |
| 1 | Vitantonio Liuzzi | Mark Webber |
| 0 | Scott Speed | Nico Rosberg |
| 0 | Christijan Albers | Christijan Albers |
| 0 | Tiago Monteiro | Tiago Monteiro |
| 0 | Takuma Sato | Takuma Sato |

Table 3.1.: The official ranking of the Formula One season 2006 and the Kemeny consensus.

#### 3.2.2.2. Winter sports

The properties of the Formula One instances are not very fortunate for the running times of our search tree algorithms. Especially the high rate of dirty pairs seems to be hard as we have already seen for the randomized data. Now, we want to investigate whether this holds for other sports competitions. Therefore, we created three further instances based on different winter sports competitions. One is generated by using the cross-skiing (15 km men) competitions of the season 2008/2009. For another instance we use biathlon team results of the season 2008/2009. And for the third we use the overall results of the seasons 2006-2009 in cross-skiing championship and rank the best 75 sportsmen of each season.

**Properties** We got instances ranging from 10 to 23 candidates. In contrast to the Formula One elections, all three instances have low rates of dirty pairs (about 50-75%) and higher rates of majority-non-dirty pairs (about 60-80%). The data reduction rule “Condorcet winner/looser” could not remove any candidates.

**Results** We computed the Kemeny score efficiently with `kconsens_cands`, `kconsens_triples`, or `4-kconsens`. All instances could be solved in at most few hours. So, we found instances that are generated from sports competitions where the search tree algorithms are efficient. Together with the results from the Formula One instances one can summarize that it seems to depend on the concrete sports competition which is the best algorithm. More detailed information can be found in the Appendix.

### 3.2.3. Search engines

Generating ranking data based on web search results can be realized with different methods. The first method is very intuitive: We define a search term and query several search engines. In our case we use the popular search engines google, yahoo, ask, and bing (formerly known as msn live search). Each search result provides a preference list of web-links. It is reasonable to remove web-links, that only appear in one single search result. It it realistic to assume that such search web-links are not of particular interest. The significance and size of the generated election highly depends on the search term. In some cases promising candidates will be removed, because different search engines return urls, that variate a little, but result in the same website. Therefore several filters are helpful to produce more interesting elections. To demonstrate this, we generated another set of instances, where every url is reduced to its domain. The generation method can be easily extended if we customize the search parameters of the engines. One successful example was to request one search term in different languages. Here it is also possible (and sometimes necessary) to translate the search term. We use the same search terms as used in [SvZ09]. The second method is based on the semantic similarity of different search terms. We define a list of search terms and query the same search engine. Each search result provides a preference list of web-links again. Here, it is very important that the search terms are really semantically similar. Otherwise, we have to request too many results for each search term to find congruent urls. The last method of generating ranking data from search results breaks with the idea of meta search engines. We define a list of search terms, each corresponds to one candidate. We generate the preference lists by requesting the search term of a specific search engine, one for each vote, and sort them according the number of search results. Thus, ties in the preference lists are possible, but improbable. An example instance is a list of some metropolises.

**Properties** Now, we consider the properties of the generated data. Although we use three different methods for the generation of the instances, the properties and results are similar for all three types. In most cases about 50% or less of the candidate pairs are dirty. Moreover, analysing the 2/3-majorities even more than 75% of the candidate pairs are non-dirty. We have also checked if we could apply the reduction rule “Condorcet winner/looser” on the instances. In ca. 50% of the cases we could delete between one and six candidates. In two cases all candidates could be removed: Searching for “Lipari” and “recycling cans” generated polynomial-time solvable special cases as discussed in Section 2.4. The maximum range was about 50-80% of the maximal possible value in the most instances. We can also see all results detailed in the appendix (Section A.3.1).

**Results** We are able to compute the optimal consensus of most instances with up to 30 candidates efficiently. For some instances we get remarkable results: Searching for “citrus groves” produced an election with eleven candidates. The algorithm `kconsens_cands` was able to compute the Kemeny score in 0.281 seconds and `4-kconsens` in 0.21 seconds. In contrast, `kconsens_cands` was substantially more efficient for all randomly generated instances with eleven candidates as well. This observation is even more clear for the search term “cheese” with a running time of more than 200 seconds for `kconsens_cands` 52.11 seconds for `kconsens_triples` and only 0.291 seconds for `4-kconsens`, for an election with 18 candidates. We can see more such instances (like “bicycling”) in the results table (Section A.3.1) in the Appendix. Another interesting observation is that the size of the dirty-sets we use in the search tree influences the running time very much in some instances: Searching for “classical guitar” produced an election, where we compute the Kemeny score with `kconsens_cands` in 112 seconds, with `4-kconsens` in 105 seconds and with `5-kconsens` in only 0.041 seconds. Otherwise we get a little bit increased running time if we use $t$-`kconsens` with $t > 5$. See Figure 3.3. We solve KEMENY SCORE with $t$-`kconsens` for $t = \{4, \ldots, 10\}$ on the elections that are produced by the web searches for “classical guitar” and “java”.

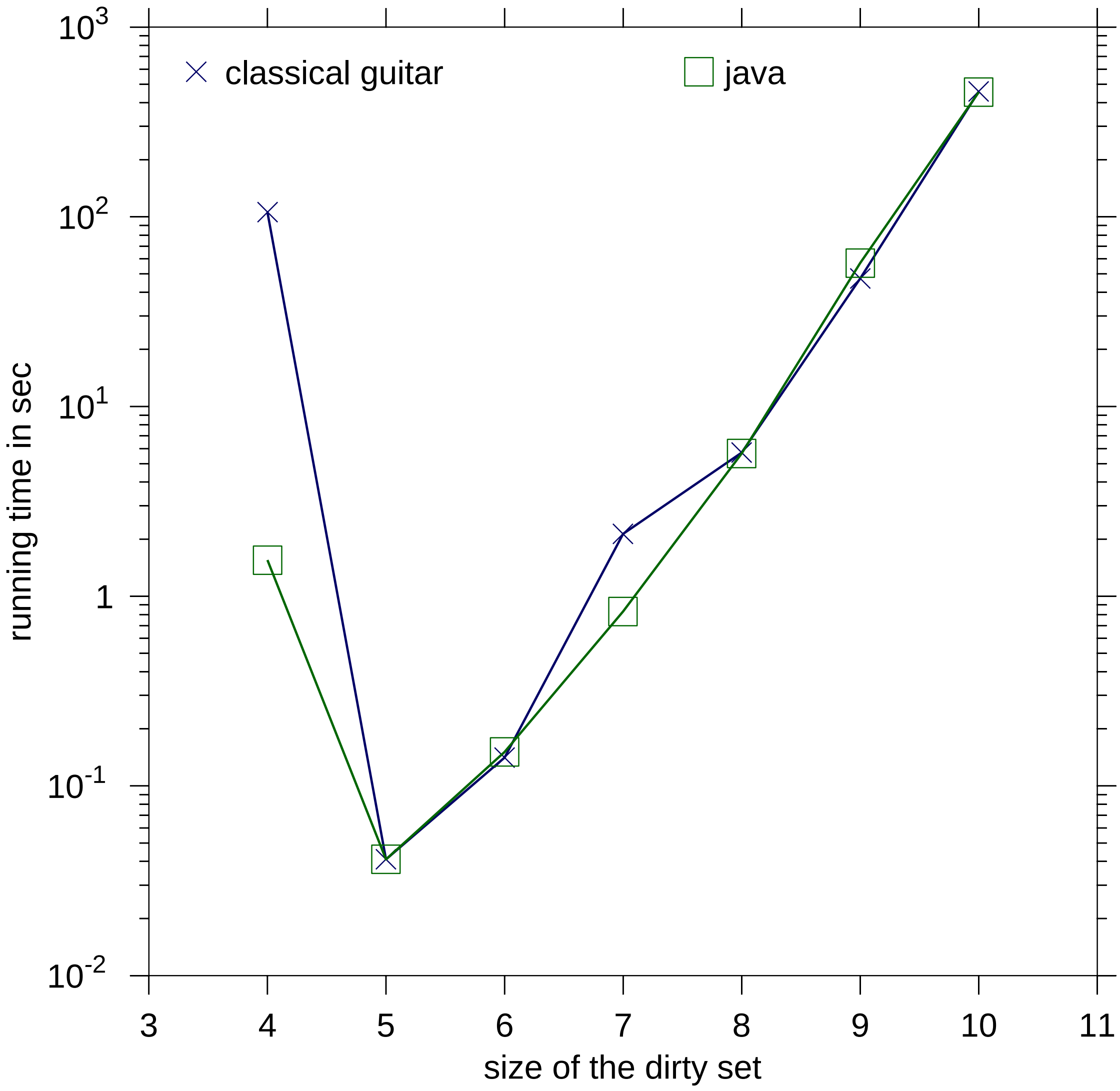

Figure 3.3.: Web search data: Running-time against the size of the dirty set.

# 4. Conclusion

This work shows once more [GN03, HKMN09, SW, BBL08], that parameterized algorithms are not only of theoretical interest. We provide implementations for two types of parameterized algorithms that solve the Kemeny Score problem. With these implementations we are able to compute an optimal Kemeny consensus for real-world data with different scopes of applications efficiently for instances of interesting sizes. We generalize a type of search tree algorithms that branches over dirty sets of candidates. We found a better upper bound for the worst-case running-time if the size of these sets is four, compared to the known results for sizes two and three. Independently from this work, an even more improved algorithm was developed in [Sim09] with a different approach. The result was a search tree algorithm with a very similar branching strategy. However, we implemented and tested a search tree algorithm for arbitrary sizes of the dirty sets and showed the upper bound for the worst-case running time only for the constant size of 4. There is hope that it is possible to find even better bounds for greater sets. Perhaps one can find a generalized upper bound for the running time depending on the size of the dirty set. In addition, we implemented a dynamic programming algorithm, which is a parameterization for the parameter "number of candidates". Particularly, the tests showed that, depending on the parameter, both algorithms can be much faster. Elections with only a few candidates (up to 30) can be solved with the dynamic programming algorithm efficiently. The search tree algorithms are much better if one has instances with only a few dirty sets or if the Kemeny score is similar to the computed lower bound, that is the minimum subscore of the set of all candidate pairs. Instances generated from web search engines seems to have this properties as well as some sports competition rankings. Furthermore, for the search tree algorithm it is not clear which size of the dirty-set is the best. The tests show different behaviour on several instances even if they are of the same type. It might be interesting to analyse this behaviour in future work. We do not know whether it is possible to compute special properties of the instances, which are indicators for the best size of the dirty sets. However, there are several other parameterizations [BFG+08] for the Kemeny Score problem. It should be informative to implement them and compare the algorithms in further experiments. Anyway, one could provide a framework of parameterized algorithms that solve Kemeny Score. For every instance the probably most efficient algorithms would be used. In some cases it might be reasonable for the framework to run several (probably fast) algorithm parallel. Beside the data reduction rule "Condorcet winner/looser" some other reduction rules for Kemeny Score were developed [BFG+08, BGKN09]. Since we could show that our real world data instanced could be reduced (partially significant), it sounds promising to develop and test further more data reduction rules.

# A. Test data

**Properties overview**

| property | short name | explanation |
|---|---|---|
| the number of votes | #votes ($n$) | |
| the number of candidates | #candidates ($m$) | |
| number of swaps | $w$ | The expected number of swaps per vote. |
| swap range | $d$ | The maximum distance of two (swapped) candidates regarding to the reference vote. |
| the number dirty pairs | #dirty pairs ($p$) | |
| rate of dirty pairs | %dirty pairs | How many pairs are dirty in percent. |
| the number majority-dirty pairs | #dirty pairs | |
| rate of majority-dirty pairs | %dirty pairs | How many pairs are majority-dirty in percent. |
| minimal score | min score | The minimal pairwise score as lower bound. |
| maximal score | max score | The maximal pairwise score as upper bound. |
| maximum range | max range | |
| average kt-distance | average kt-dist | |
| reduced candidates | red. cands. | How many candidates could be deleted by the data reduction rule. |

The pairwise score is the subscore of the set of all candidate pairs.

# A.1. Sports competitions

## A.1.1. Formula One instances

**Properties of the Formula One instances**

| season | #votes | #candi-dates | #dirty pairs | %dirty pairs | #maj. pairs | %maj. pairs | min score | max score | max range | average kt-dist | red. cands. |
|---|---|---|---|---|---|---|---|---|---|---|---|
| 1961 | 8 | 9 | 36 | 100.00 | 13 | 36.11 | 96 | 192 | 8 | 15.96 | 0 |
| 1962 | 9 | 8 | 28 | 100.00 | 8 | 28.57 | 79 | 173 | 7 | 11.81 | 0 |
| 1963 | 10 | 9 | 34 | 94.44 | 12 | 33.33 | 122 | 238 | 8 | 15.20 | 1 |
| 1964 | 10 | 9 | 36 | 100.00 | 10 | 27.78 | 143 | 217 | 8 | 17.44 | 0 |
| 1965 | 10 | 7 | 19 | 90.48 | 13 | 61.90 | 56 | 154 | 6 | 7.16 | 0 |
| 1966 | 9 | 8 | 28 | 100.00 | 7 | 25.00 | 88 | 164 | 7 | 12.58 | 0 |
| 1967 | 11 | 8 | 28 | 100.00 | 13 | 46.43 | 97 | 211 | 7 | 11.53 | 0 |
| 1968 | 12 | 8 | 28 | 100.00 | 10 | 35.71 | 111 | 225 | 7 | 12.08 | 0 |
| 1969 | 11 | 6 | 15 | 100.00 | 4 | 26.67 | 61 | 104 | 5 | 6.47 | 0 |
| 1970 | 13 | 10 | 45 | 100.00 | 15 | 33.33 | 216 | 369 | 9 | 20.85 | 0 |
| 1971 | 11 | 11 | 55 | 100.00 | 21 | 38.18 | 209 | 396 | 10 | 25.05 | 0 |
| 1972 | 12 | 11 | 55 | 100.00 | 23 | 41.82 | 204 | 456 | 10 | 23.08 | 0 |
| 1973 | 15 | 12 | 66 | 100.00 | 30 | 45.45 | 306 | 684 | 11 | 27.59 | 0 |
| 1974 | 15 | 14 | 91 | 100.00 | 35 | 38.46 | 457 | 908 | 13 | 40.35 | 0 |
| 1975 | 14 | 13 | 78 | 100.00 | 31 | 39.74 | 371 | 721 | 12 | 34.82 | 0 |
| 1976 | 16 | 13 | 78 | 100.00 | 39 | 50.00 | 410 | 838 | 12 | 33.62 | 0 |
| 1977 | 17 | 13 | 78 | 100.00 | 24 | 30.77 | 475 | 851 | 12 | 35.74 | 0 |
| 1978 | 16 | 16 | 117 | 97.50 | 69 | 57.50 | 597 | 1323 | 15 | 49.77 | 0 |
| 1979 | 15 | 19 | 168 | 98.25 | 61 | 35.67 | 823 | 1742 | 18 | 73.13 | 0 |
| 1980 | 14 | 19 | 164 | 95.91 | 91 | 53.22 | 712 | 1682 | 17 | 69.09 | 1 |
| 1981 | 15 | 19 | 167 | 97.66 | 80 | 46.78 | 767 | 1798 | 18 | 69.44 | 1 |
| 1982 | 16 | 9 | 35 | 97.22 | 21 | 58.33 | 178 | 398 | 8 | 14.33 | 0 |
| 1983 | 15 | 24 | 273 | 98.91 | 117 | 42.39 | 1282 | 2858 | 23 | 116.88 | 0 |
| 1984 | 16 | 19 | 170 | 99.42 | 89 | 52.05 | 886 | 1850 | 18 | 74.28 | 0 |
| 1985 | 16 | 14 | 91 | 100.00 | 53 | 58.24 | 458 | 998 | 13 | 38.45 | 0 |
| 1986 | 16 | 21 | 207 | 98.57 | 136 | 64.76 | 965 | 2395 | 20 | 83.89 | 0 |
| 1987 | 16 | 21 | 209 | 99.52 | 121 | 57.62 | 1026 | 2334 | 20 | 87.08 | 0 |
| 1988 | 16 | 28 | 357 | 94.44 | 252 | 66.67 | 1568 | 4480 | 24 | 137.85 | 0 |
| 1989 | 16 | 26 | 289 | 88.92 | 223 | 68.62 | 1285 | 3915 | 24 | 111.34 | 0 |
| 1990 | 16 | 24 | 249 | 90.22 | 194 | 70.29 | 1090 | 3326 | 22 | 96.28 | 0 |
| 1991 | 16 | 24 | 262 | 94.93 | 173 | 62.68 | 1178 | 3238 | 21 | 100.60 | 0 |
| 1992 | 16 | 22 | 229 | 99.13 | 130 | 56.28 | 1141 | 2555 | 21 | 97.60 | 1 |
| 1993 | 16 | 18 | 151 | 98.69 | 79 | 51.63 | 775 | 1673 | 17 | 64.68 | 0 |
| 1994 | 16 | 16 | 113 | 94.17 | 62 | 51.67 | 558 | 1362 | 15 | 45.33 | 0 |
| 1995 | 17 | 16 | 120 | 100.00 | 72 | 60.00 | 611 | 1429 | 15 | 49.40 | 0 |
| 1996 | 16 | 19 | 171 | 100.00 | 106 | 61.99 | 834 | 1902 | 18 | 71.78 | 0 |
| 1997 | 17 | 18 | 153 | 100.00 | 80 | 52.29 | 849 | 1752 | 17 | 67.49 | 0 |
| 1998 | 16 | 21 | 206 | 98.10 | 146 | 69.52 | 889 | 2471 | 20 | 78.53 | 0 |
| 1999 | 16 | 19 | 167 | 97.66 | 93 | 54.39 | 847 | 1889 | 18 | 70.19 | 0 |
| 2000 | 17 | 22 | 230 | 99.57 | 124 | 53.68 | 1170 | 2757 | 21 | 94.29 | 0 |
| 2001 | 17 | 18 | 152 | 99.35 | 67 | 43.79 | 819 | 1782 | 17 | 64.22 | 1 |
| 2002 | 17 | 18 | 140 | 91.50 | 88 | 57.52 | 751 | 1850 | 17 | 59.68 | 1 |
| 2003 | 16 | 16 | 118 | 98.33 | 79 | 65.83 | 583 | 1337 | 15 | 49.47 | 0 |
| 2004 | 18 | 15 | 101 | 96.19 | 73 | 69.52 | 425 | 1465 | 14 | 33.36 | 2 |
| 2005 | 19 | 13 | 78 | 100.00 | 58 | 74.36 | 394 | 1088 | 12 | 29.05 | 0 |
| 2006 | 18 | 18 | 152 | 99.35 | 100 | 65.36 | 682 | 2072 | 17 | 54.41 | 0 |
| 2007 | 17 | 18 | 149 | 97.39 | 108 | 70.59 | 602 | 1999 | 17 | 49.91 | 0 |
| 2008 | 18 | 20 | 182 | 95.79 | 111 | 58.42 | 923 | 2497 | 19 | 71.47 | 0 |

**Running times (in seconds) for the Formula One instances**

| season | kconsens_cands | kconsens_pairs | kconsens_triples | 4-kconsens | 5-kconsens | 6-kconsens |
|---|---|---|---|---|---|---|
| 1961 | 0.051 | n/a | 108.461 | 32.031 | 59.121 | 147.581 |
| 1962 | 0.021 | n/a | 7.371 | 2.221 | 4.281 | 11.811 |
| 1963 | 0.041 | n/a | 68.701 | 25.711 | 48.771 | 111.531 |
| 1964 | 0.051 | n/a | n/a | 68.721 | 147.401 | n/a |
| 1965 | 0.001 | 0.651 | 0.021 | 0.001 | 0.011 | 0.021 |
| 1966 | 0.021 | n/a | 13.611 | 3.071 | 6.151 | 18.761 |
| 1967 | 0.021 | n/a | 4.201 | 0.981 | 2.231 | 5.851 |
| 1968 | 0.021 | n/a | 1.431 | 0.001 | 0.011 | 0.061 |
| 1969 | 0.001 | 0.241 | 0.111 | 0.051 | 0.071 | 0.091 |
| 1970 | 0.111 | n/a | n/a | 705.771 | n/a | n/a |
| 1971 | 0.281 | n/a | n/a | 5831.82 | n/a | n/a |
| 1972 | 0.281 | n/a | n/a | 2042 | n/a | n/a |
| 1973 | 0.861 | n/a | n/a | n/a | n/a | n/a |
| 1974 | 11.351 | n/a | n/a | n/a | n/a | n/a |
| 1975 | 2.931 | n/a | n/a | n/a | n/a | n/a |
| 1976 | 2.921 | n/a | n/a | n/a | n/a | n/a |
| 1977 | 2.931 | n/a | n/a | n/a | n/a | n/a |
| 1978 | 111.961 | n/a | n/a | n/a | n/a | 0.241 |
| 1979 | 8426.59 | n/a | n/a | n/a | n/a | n/a |
| 1980 | 1927.58 | n/a | n/a | n/a | n/a | n/a |
| 1981 | 1992.62 | n/a | n/a | n/a | n/a | n/a |
| 1982 | 0.051 | n/a | 4.811 | 0.001 | 0.631 | 0.771 |
| 1983 | a few hours | n/a | n/a | n/a | n/a | n/a |
| 1984 | 8310.57 | n/a | n/a | n/a | n/a | n/a |
| 1985 | 10.571 | n/a | n/a | n/a | n/a | n/a |
| 1986 | a few hours | n/a | n/a | n/a | n/a | n/a |
| 1987 | a few hours | n/a | n/a | n/a | n/a | n/a |
| 1988 | a few days | n/a | n/a | n/a | n/a | n/a |
| 1989 | a few days | n/a | n/a | n/a | n/a | n/a |
| 1990 | a few days | n/a | n/a | n/a | n/a | n/a |
| 1991 | a few days | n/a | n/a | n/a | n/a | n/a |
| 1992 | a few hours | n/a | n/a | n/a | n/a | n/a |
| 1993 | 1962.81 | n/a | n/a | n/a | n/a | n/a |
| 1994 | 469.821 | n/a | n/a | n/a | n/a | n/a |
| 1995 | 101.911 | n/a | n/a | n/a | n/a | n/a |
| 1996 | 8544.33 | n/a | n/a | n/a | n/a | n/a |
| 1997 | 1966.39 | n/a | n/a | n/a | n/a | n/a |
| 1998 | a few hours | n/a | n/a | n/a | n/a | n/a |
| 1999 | 8279.19 | n/a | n/a | n/a | n/a | n/a |
| 2000 | a few hours | n/a | n/a | n/a | n/a | n/a |
| 2001 | 494.791 | n/a | n/a | n/a | n/a | n/a |
| 2002 | 484.911 | n/a | n/a | n/a | n/a | n/a |
| 2003 | 114.471 | n/a | n/a | n/a | n/a | 0.201 |
| 2004 | 49.161 | n/a | n/a | n/a | n/a | 0.191 |
| 2005 | 2.821 | n/a | n/a | n/a | n/a | n/a |
| 2006 | 1929.25 | n/a | n/a | n/a | n/a | n/a |
| 2007 | 2015.77 | n/a | n/a | n/a | n/a | n/a |
| 2008 | a few hours | n/a | n/a | n/a | n/a | n/a |

Time values are not available for the search tree algorithms if test runs took more than three hours.

## A.1.2. Winter sports instances

**Properties of the winter sports instances**

| competition | #votes | #candi-dates | #dirty pairs | %dirty pairs | #maj. pairs | %maj. pairs | min score | max score | max range | average kt-dist | red. cands. |
|---|---|---|---|---|---|---|---|---|---|---|---|
| biathlon team men 08/09 | 6 | 15 | 67 | 63.81 | 65 | 61.90 | 124 | 506 | 14 | 30.60 | 0 |
| cross skiing 15km men 08/09 | 4 | 10 | 24 | 53.33 | 37 | 82.22 | 32 | 148 | 8 | 12.33 | 0 |
| cross skiing seasons 06-09 | 4 | 23 | 192 | 75.89 | 190 | 75.10 | 255 | 757 | 19 | 107.17 | 0 |

**Running times (in seconds) for the winter sports instances**

| competition | `kconsens_cands` | `kconsens_pairs` | `kconsens_triples` | 4-`kconsens` | 5-`kconsens` | 6-`kconsens` |
|---|---|---|---|---|---|---|
| biathlon team men 08/09 | 58.23 | n/a | n/a | 16.47 | 4904 | n/a |
| cross skiing 15km men 08/09 | 0.081 | 30.311 | 0.721 | 0.001 | 0.011 | 0.021 |
| cross skiing seasons 06-09 | a few hours | n/a | n/a | n/a | n/a | n/a |

Time values are not available for the search tree algorithms if test runs took more than two hours.

# A.2. Randomly generated instances

## A.2.1. Parameter values and running times for randomly generated instances

### Test series 1

We generated 10 instances with 200 votes for each parameter set. The running-times are average running-times.

**Running times and properties for test series 1**

| $m = w = d$ | 4 | 5 | 6 | 7 | 8 | 9 | 10 | 11 | 12 | 13 | 14 | 15 | 16 | 17 |
|---|---|---|---|---|---|---|---|---|---|---|---|---|---|---|
| kconsens_cands | 0.001 | 0.002 | 0.006 | 0.015 | 0.025 | 0.058 | 0.117 | 0.302 | 0.891 | 2.89 | 11.1 | 48.8 | 113 | 488 |
| kconsens_pairs | 0.001 | 0.013 | 0.469 | 43.21 | - | - | - | - | - | - | - | - | - | - |
| kconsens_triples | 0.003 | 0.017 | 0.0346 | 3.822 | 132.98 | - | - | - | - | - | - | - | - | - |
| 4-kconsens | 0.002 | 0.010 | 0.131 | 1.81 | 37.73 | 546.38 | - | - | - | - | - | - | - | - |
| # dirty pairs | 6 | 10 | 15 | 21 | 28 | 36 | 45 | 55 | 66 | 78 | 91 | 105 | 120 | 126 |

### Test series 2

We generated instances with 14 candidates. The running times are average running-times.

**Running times (in seconds) and properties for test series 2**

| $n$ | $p$ | $d$ | $w$ | kconsens_cands | kconsens_pairs | kconsens_triples | 4-kconsens |
|---|---|---|---|---|---|---|---|
| 5 | 5 | 2 | 4 | 17.071 | 0.001 | 0.001 | 0.001 |
| 10 | 11 | 2 | 7 | 17.211 | 0.041 | 0.021 | 0.011 |
| 16 | 13 | 2 | 11 | 17.501 | 0.171 | 0.031 | 0.011 |
| 10 | 26 | 4 | 7 | 17.611 | n/a | 0.061 | 0.021 |
| 16 | 29 | 4 | 11 | 17.201 | n/a | 0.981 | 0.021 |
| 22 | 34 | 4 | 15 | 17.221 | n/a | 1.601 | 0.021 |
| 10 | 43 | 6 | 7 | 17.071 | n/a | 6.311 | 0.031 |
| 22 | 45 | 6 | 15 | 13.271 | n/a | 8.761 | 0.031 |
| 10 | 52 | 8 | 7 | 17.561 | n/a | n/a | 0.041 |
| 10 | 53 | 10 | 7 | 17.621 | n/a | n/a | 0.061 |
| 16 | 58 | 8 | 11 | 17.561 | n/a | n/a | 0.061 |
| 22 | 59 | 8 | 15 | 17.161 | n/a | n/a | n/a |
| 10 | 65 | 12 | 7 | 17.181 | n/a | n/a | n/a |
| 22 | 69 | 10 | 15 | 13.271 | n/a | n/a | n/a |
| 10 | 74 | 14 | 7 | 17.521 | n/a | n/a | n/a |
| 16 | 75 | 12 | 11 | 17.081 | n/a | n/a | n/a |
| 16 | 80 | 14 | 11 | 13.351 | n/a | n/a | n/a |
| 22 | 86 | 12 | 15 | 13.281 | n/a | n/a | n/a |

Time values are not available for the search tree algorithms if test runs took more than 30 minutes.

# A.3. Search engines

## A.3.1. Attributes of the websearch instances

Websearches with the search engines google, yahoo, ask, and msn with 300 search results per engine.

**Properties of the websearch instances**

| search term | #candi-dates | #dirty pairs | %dirty pairs | #maj. pairs | %maj. pairs | min score | max score | max range | average kt-dist | red. cands. |
|---|---|---|---|---|---|---|---|---|---|---|
| affirmative+action | 22 | 101 | 43.72 | 203 | 87.88 | 129 | 795 | 16 | 54.33 | 6 |
| alcoholism | 14 | 33 | 36.26 | 76 | 83.52 | 48 | 316 | 8 | 18.00 | 0 |
| amusement+parks | 11 | 31 | 56.36 | 44 | 80.00 | 42 | 178 | 9 | 16.83 | 1 |
| architecture | 17 | 53 | 38.97 | 121 | 88.97 | 68 | 476 | 11 | 27.50 | 0 |
| bicycling | 20 | 85 | 44.74 | 162 | 85.26 | 113 | 647 | 15 | 45.83 | 4 |
| blues | 20 | 77 | 40.53 | 158 | 83.16 | 109 | 651 | 12 | 43.00 | 3 |
| cheese | 18 | 74 | 48.37 | 131 | 85.62 | 96 | 516 | 13 | 39.83 | 3 |
| citrus+groves | 11 | 39 | 70.91 | 46 | 83.64 | 48 | 172 | 9 | 20.17 | 1 |
| classical+guitar | 16 | 62 | 51.67 | 105 | 87.50 | 77 | 403 | 12 | 32.83 | 1 |
| computer+vision | 18 | 115 | 75.16 | 116 | 75.82 | 152 | 460 | 15 | 62.83 | 1 |
| cruises | 18 | 112 | 73.20 | 113 | 73.86 | 152 | 460 | 14 | 62.17 | 1 |
| Death+Valley | 17 | 63 | 46.32 | 120 | 88.24 | 79 | 465 | 13 | 33.17 | 1 |
| field+hockey | 18 | 63 | 41.18 | 130 | 84.97 | 86 | 526 | 11 | 34.50 | 0 |
| gardening | 17 | 49 | 36.03 | 123 | 90.44 | 62 | 482 | 10 | 26.17 | 2 |
| graphic+design | 11 | 21 | 38.18 | 49 | 89.09 | 27 | 193 | 6 | 10.67 | 0 |
| Gulf+war | 18 | 62 | 40.52 | 129 | 84.31 | 86 | 526 | 13 | 34.17 | 2 |
| HIV | 15 | 44 | 41.90 | 93 | 88.57 | 56 | 364 | 11 | 23.00 | 3 |
| java | 17 | 65 | 47.79 | 119 | 87.50 | 82 | 462 | 14 | 34.50 | 4 |
| Lipari | 6 | 4 | 26.67 | 15 | 100.00 | 4 | 56 | 3 | 1.50 | 6 |
| lyme+disease | 20 | 81 | 42.63 | 172 | 90.53 | 99 | 661 | 13 | 42.00 | 0 |
| mutual+funds | 13 | 30 | 38.46 | 69 | 88.46 | 39 | 273 | 7 | 15.50 | 1 |
| National+parks | 13 | 32 | 41.03 | 73 | 93.59 | 37 | 275 | 10 | 15.67 | 6 |
| parallel+architecture | 6 | 8 | 53.33 | 12 | 80.00 | 11 | 49 | 4 | 3.67 | 2 |
| Penelope+Fitzgerald | 14 | 65 | 71.43 | 65 | 71.43 | 91 | 273 | 12 | 36.33 | 1 |
| recycling+cans | 3 | 1 | 33.33 | 3 | 100.00 | 1 | 11 | 2 | 0.50 | 3 |
| rock+climbing | 15 | 84 | 80.00 | 76 | 72.38 | 113 | 307 | 14 | 46.00 | 1 |
| San+Francisco | 6 | 9 | 53.33 | 11 | 73.33 | 12 | 48 | 4 | 4.17 | 1 |
| Shakespeare | 26 | 175 | 53.85 | 265 | 81.54 | 235 | 1065 | 21 | 96.17 | 2 |
| stamp+collection | 9 | 19 | 52.78 | 31 | 86.11 | 24 | 120 | 8 | 9.17 | 3 |
| sushi | 14 | 40 | 43.96 | 77 | 84.62 | 54 | 310 | 10 | 21.83 | 2 |
| table+tennis | 13 | 38 | 48.72 | 68 | 87.18 | 48 | 264 | 9 | 19.83 | 2 |
| telecommuting | 12 | 28 | 42.42 | 53 | 80.30 | 41 | 223 | 7 | 15.17 | 2 |
| Thailand+tourism | 16 | 57 | 47.50 | 101 | 84.17 | 76 | 404 | 11 | 30.67 | 1 |
| vintage+cars | 4 | 2 | 33.33 | 5 | 83.33 | 3 | 21 | 2 | 0.50 | 2 |
| volcano | 20 | 98 | 51.58 | 159 | 83.68 | 129 | 631 | 17 | 53.17 | 0 |
| zen+budism | 11 | 22 | 40.00 | 49 | 89.09 | 28 | 192 | 7 | 11.17 | 1 |
| Zener | 14 | 54 | 59.34 | 74 | 81.32 | 71 | 293 | 13 | 29.00 | 1 |

**Running times (in seconds) for the websearch instances**

| search term | kconsens_cands | kconsens_pairs | kconsens_triples | 4-kconsens | 5-kconsens | 6-kconsens |
|---|---|---|---|---|---|---|
| affirmative+action | 114.931 | n/a | n/a | n/a | 0.051 | 0.211 |
| alcoholism | 10.871 | n/a | 4.161 | 0.011 | 0.021 | 0.101 |
| amusement+parks | 0.081 | 34.501 | 0.001 | 0.011 | 0.001 | 0.041 |
| architecture | 475.921 | n/a | n/a | 0.021 | 21.871 | 0.141 |
| bicycling | 109.101 | n/a | n/a | n/a | 0.101 | n/a |
| blues | 486.071 | n/a | n/a | n/a | 2.941 | n/a |
| cheese | 49.891 | n/a | n/a | 1724.79 | 0.041 | 0.131 |
| citrus+groves | 0.081 | n/a | 6.591 | 0.001 | 0.011 | 0.061 |
| classical+guitar | 50.741 | n/a | n/a | 1087.12 | 0.031 | 0.281 |
| computer+vision | 486.201 | n/a | n/a | n/a | n/a | n/a |
| cruises | 475.591 | n/a | n/a | n/a | n/a | n/a |
| Death+Valley | 109.531 | n/a | n/a | 16.901 | 0.041 | 0.151 |
| field+hockey | 2060.72 | n/a | n/a | n/a | n/a | n/a |
| gardening | 53.381 | n/a | n/a | 0.021 | 0.031 | 0.101 |
| graphic+design | 0.261 | 15.611 | 0.081 | 0.001 | 0.011 | 0.031 |
| Gulf+war | 117.931 | n/a | n/a | 126.071 | 0.041 | 3.021 |
| HIV | 0.811 | n/a | 7.131 | 0.001 | 0.011 | 0.061 |
| java | 2.821 | n/a | n/a | 73.441 | 0.021 | 0.361 |
| Lipari | 0.001 | 0.001 | 0.001 | n/a | n/a | n/a |
| lyme+disease | a few hours | n/a | n/a | n/a | n/a | 0.221 |
| mutual+funds | 0.791 | n/a | 0.851 | 0.001 | 0.011 | 0.041 |
| National+parks | 0.001 | 0.041 | 0.011 | 0.001 | 0.001 | 0.011 |
| parallel+architecture | 0.001 | 0.001 | 0.001 | 0.001 | 0.001 | 0.001 |
| Penelope+Fitzgerald | 2.841 | n/a | n/a | 0.031 | 136.241 | n/a |
| recycling+cans | 0.001 | 0.001 | 0.001 | n/a | n/a | n/a |
| rock+climbing | 11.831 | n/a | n/a | n/a | n/a | n/a |
| San+Francisco | 0.001 | 0.001 | 0.011 | 0.001 | 0.011 | 0.011 |
| Shakespeare | a few hours | n/a | n/a | n/a | n/a | n/a |
| stamp+collection | 0.001 | 0.051 | 0.011 | 0.001 | 0.001 | 0.011 |
| sushi | 0.801 | n/a | 32.461 | 0.201 | 0.221 | 0.061 |
| table+tennis | 0.261 | n/a | 0.301 | 0.001 | 0.021 | 0.061 |
| telecommuting | 0.091 | 0.261 | 0.021 | 0.001 | 0.011 | 0.041 |
| Thailand+tourism | 51.261 | n/a | n/a | 4407.3 | 4109.55 | n/a |
| vintage+cars | 0.001 | 0.001 | 0.001 | 0.001 | 0.001 | 0.001 |
| volcano | a few hours | n/a | n/a | n/a | n/a | n/a |
| zen+budism | 0.081 | 35.571 | 0.291 | 0.001 | 0.001 | 0.041 |
| Zener | 2.791 | n/a | n/a | 0.011 | 0.041 | 0.101 |

Time values are not available for the search tree algorithms if test runs took more than three hours.

Websearches with the search terms: “hotel dublin” “hotels dublin” “rooms dublin” “bed dublin” with 300 search results per engine. Result urls are reduced to the domain to get greater instances.

**Properties**

| search-engine | #candi-dates | #dirty pairs | %dirty pairs | #maj. pairs | %maj. pairs | min score | max score | max range | average kt-dist | red. cands. |
|---|---|---|---|---|---|---|---|---|---|---|
| ask | 4 | 4 | 66.67 | 5 | 83.33 | 5 | 19 | 3 | 2.17 | 2 |
| google | 13 | 60 | 76.92 | 53 | 67.95 | 85 | 227 | 11 | 33.00 | 0 |
| msnlive | 13 | 14 | 17.95 | 73 | 93.59 | 19 | 293 | 6 | 7.17 | 6 |
| yahoo | 12 | 49 | 74.24 | 50 | 75.76 | 65 | 199 | 11 | 26.50 | 2 |

**Running times (in seconds)**

| search engine | `kconsens_cands` | `kconsens_pairs` | `kconsens_triples` | `4-kconsens` | `5-kconsens` | `6-kconsens` |
|---|---|---|---|---|---|---|
| ask | 0.001 | n/a | n/a | 0.001 | 0.001 | 0.001 |
| google | 2.431 | n/a | n/a | 0.051 | 0.061 | 0.101 |
| msnlive | 0.001 | n/a | n/a | 0.001 | 0.001 | 0.021 |
| yahoo | 0.081 | n/a | n/a | 8.121 | 5.321 | 0.121 |

Time values are not available for the search tree algorithms if test runs took more than 30 minutes.

Websearches with the search term list: Paris, London, Washington, Madrid, Berlin, Ottawa, Wien, Canberra, Peking, Prag, Moskau. We generated one vote for each of the search engines: google, yahoo, ask, and msn. The candidates are ranked according to the corresponding number of result pages.

**Properties**

| #votes | #candi-dates | #dirty pairs | %dirty pairs | #maj. pairs | %maj. pairs | min score | max score | max range | average kt-dist | red. cands. |
|---|---|---|---|---|---|---|---|---|---|---|
| 4 | 11 | 19 | 34.55 | 46 | 83.64 | 28 | 192 | 6 | 10.33 | 0 |

**Running times (in seconds)**

| `kconsens_cands` | `kconsens_pairs` | `kconsens_triples` | `4-kconsens` | `5-kconsens` | `6-kconsens` |
|---|---|---|---|---|---|
| 0.231 | 3.501 | 0.001 | 0.001 | 0.001 | 0.0041 |